\documentclass[a4paper,10pt]{article} %

\usepackage[authoryear]{natbib}
\usepackage{graphicx} %
\usepackage{bm}
\usepackage{mathtools}
\usepackage[hidelinks]{hyperref}
\usepackage{xcolor}
\usepackage{xargs}
\usepackage{amsmath, amssymb, amsthm, cleveref}
\usepackage{subfig}
\usepackage{mathrsfs}
\usepackage{fullpage}

\newtheorem{remark}{Remark}
\DeclareMathOperator{\diver}{div}
\DeclareMathOperator{\grad}{grad}
\DeclareMathOperator{\tr}{tr}
\let\det\relax
\DeclareMathOperator{\det}{det}
\newcommand{\vect}[1]{\boldsymbol{#1}}
\newcommand{\tens}[1]{\mathsf{#1}}

\begin{document}
\title{\textsc{Elastic Plateau--Rayleigh instability in soft cylinders: Surface elasticity and periodic beading}}

\author{\textsc{F. Magni}\thanks{\href{mailto:fmagni@sissa.it}{\texttt{fmagni@sissa.it}}} $\,\,\cdot$
  \textsc{D. Riccobelli}\thanks{\href{mailto:davide.riccobelli@sissa.it}{\texttt{davide.riccobelli@sissa.it}}}\bigskip\\
  \normalsize Mathematics Area, mathLab, SISSA -- International School for Advanced Studies}
\date{\today}

\maketitle

\begin{abstract}
The Plateau--Rayleigh instability shows that a cylindrical fluid flow can be destabilized by surface tension. Similarly, capillary forces can make an elastic cylinder unstable when the elastocapillary length is comparable to the cylinder's radius. 
While existing models predict a single isolated bulge as the result of an instability, experiments reveal a periodic sequence of bulges spaced out by thinned regions, a phenomenon known as beading instability. Most models assume that surface tension is independent of the deformation of the solid, neglecting variations due to surface stretch.

In this work, we assume that surface tension arises from the deformation of material particles near the free surface, treating it as a pre-stretched elastic surface surrounding the body. Using the theoretical framework proposed by Gurtin and Murdoch, we show that a cylindrical solid can undergo a mechanical instability with a finite critical wavelength if the body is sufficiently soft or axially stretched. Post-buckling numerical simulations reveal a morphology in qualitative agreement with experimental observations. Period-halving secondary bifurcations are also observed. The results of this research have broad implications for soft materials, biomechanics, and microfabrication applications where surface tension plays a crucial role.
\end{abstract}

\section{Introduction}
A common observation, such as when tap water is gently opened, is that a
thin, cylindrical stream of fluid can undergo a hydrodynamic instability. This
instability causes the formation of sinusoidal bulges along the stream, which
eventually break apart into droplets. This phenomenon is known as \emph{Plateau--Rayleigh} instability: fluid-air surface tension tends to minimize the surface area, which leads to the formation of droplets \citep{Plateau,rayleigh_xvi_1892}.

Similarly to fluids, solids also possess a surface tension at the interface with other materials. While, for most solids, surface energy is negligible compared to the elastic energy of the object, when the material is very soft or small, capillarity becomes significant and can deform elastic solid bodies \citep{style2017elastocapillarity,bico2018elastocapillarity}.
In particular, surface tension can induce large elastic deformations in soft solids, such as hydrogels filaments \citep{Mora_2013,Ang_2021}, rubber-like materials \citep{py2007capillary,elettro2016drop}, and even biological matter \citep{riccobelli2020surface,yadav2022gradients,ang2024elastocapillary,riccobelli2024elastocapillarity}.

In recent years, several studies have investigated the classical counterparts of fluid-dynamical instabilities in solids, like the elastic Rayleigh-Taylor \citep{Robinson_1989,Plohr_1998,Piriz_2009,Mora_2014,Riccobelli_2017} and Faraday \citep{Shao_2018,Bevilacqua_2020,Shao_2020}.%

Similarly, experiments show that elastic filaments can undergo a surface-tension-driven instability, resulting in a periodic sequence of bulges along the elastic body \citep{matsuo1992patterns,zuo2005experimental,naraghi2007mechanical,PhysRevLett.105.214301}, a phenomenon known as \emph{beading}. This elastic analogue of the Plateau–Rayleigh instability has been investigated in recent years, see, for instance, \cite{PhysRevLett.105.214301,Ciarletta_2012,taffetani2015elastocapillarity,xuan2016finite}, but existing theoretical models predict only the formation of an isolated bulge \citep{lestringant2020one,fu2021necking}. 

Several studies have suggested that the experimentally observed finite wavelength may result from the sensitivity of the system to imperfections; see, for example, \cite{taffetani_ciarletta, Giudici_2020}. However, the experiments by Matsuo and Tanaka \citep{matsuo1992patterns} demonstrate that the wavelength consistently scales with the radius of the cylinder. This scaling behavior indicates that wavelength selection is a robust and intrinsic feature of the instability, rather than a consequence of random surface imperfections.
The occurrence of finite wavelength selection appears to be a common feature across several experiments, as reviewed in the introduction of \cite{fu2021necking}.

\begin{figure}
    \centering
    \includegraphics[width=0.7\linewidth]{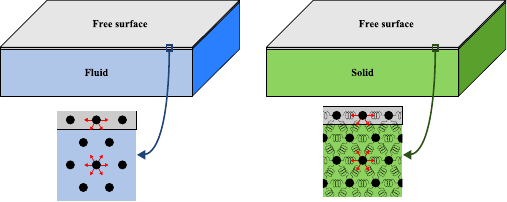}
    \caption{Surface tension in solids and fluids exhibits different constitutive responses. In fluids, it remains constant due to intermolecular forces, whereas in solids, deformation alters surface stress due to the elastic nature of intermolecular interactions.}
    \label{fig:tensione_sup_solidi_liquidi}
\end{figure}

In most of the previously cited papers, surface tension in solids is treated as a constant quantity,
similarly to fluids \citep{PhysRevLett.105.214301}. Although this might be an acceptable approximation in many
cases, studies have shown that surface tension in solids has an elastic
nature, i.e. its value depends on the deformation of the surface, see \cref{fig:tensione_sup_solidi_liquidi}. The pioneering idea behind this phenomenon is due to \cite{shuttleworth1950surface}, who postulated a linear dependence of the surface stress with respect to the surface deformation.

The inclusion of nonlinear elasticity effect in surface stress has been recently investigated in the context of the Plateau--Rayleigh instability by \cite{bakiler2023surface} and \cite{Yu_Fu}, who proposed an additive decomposition of the surface stress into a constant term, equal to the classical surface tension in a fluid, and an elastic contribution. However, the resulting critical buckling mode still corresponds to an isolated bulge \citep{Yu_Fu}.

In this work, we present a different approach, modelling surface tension in solids
as a pre-stretched elastic surface. Indeed, the rheology of fluids and solids is drastically different: particles close to the free
surface are stretched by intermolecular cohesion forces, similarly to what happens
in fluids, but this causes an elastic distortion of the material. Given the elastic nature of the body, this
phenomenon is equivalent to imposing a pre-stretch on this thin layer of particles close to the surface.
In the following, we explore how this different approach applies to the stability of cylindrical soft solids. Specifically, in Section \ref{sec:prestretched_elastic_surfaces} we propose a theoretical framework for pre-stretched elastic surfaces, building upon the work of \cite{Gurtin_Murdoch}. The proposed framework is specialized for a cylindrical geometry in Section~\ref{subsec:model}. The linear stability of the system is analysed in Section~\ref{sec:analysis_of_the_model} and a numerical post-buckling analysis is performed in Section~\ref{sec:numerical_simulations}.

\section{Pre-stretched elastic surface surrounding an elastic solid}
\label{sec:prestretched_elastic_surfaces} In this section, we present a
mathematical theory of a three-dimensional elastic solid surrounded by an
elastic surface subject to pre-stretch. The model is derived from the theoretical
framework proposed by \cite{Gurtin_Murdoch} and later extended by \cite{holland} to account for morphoelastic phenomena.

\subsection{Notation and basic kinematics}
We consider a body with reference configuration $\mathscr{B}_{0}$ surrounded
by a material surface $\mathcal{S}_{0}=\partial\mathscr{B}_{0}$. Let
$\vect{\chi}$ be the deformation field mapping $\mathscr{B}_{0}$ to the
current configuration $\mathscr{B}$. Similarly, the material reference
surface $\mathcal{S}_{0}$ is mapped to its deformed counterpart
$\mathcal{S}$ via the surface deformation map $\vect{\chi}_{s}$, which represents
the restriction of $\vect{\chi}$ to the surface.
We denote by $\mathcal{T}_{\vect{X}}$ and $\mathcal{T}_{\vect{x}}$ the tangent
spaces to $\mathcal{S}_{0}$ in $\vect{X}$ and to $\mathcal{S}$ in $\vect{x}$
respectively. %
Let $\vect{N}(\vect{X})$ and $\vect{n}(\vect{x})$ denote the outward normal
of $\mathcal{S}_{0}$ in $\vect{X}$ and of $\mathcal{S}$ in $\vect{x}$, respectively.
We introduce the surface identity tensors
\begin{align}
    \tens{I}_{s}\left(\vect{X}\right) & =\tens{I}\left(\vect{X}\right)-\vect{N}\left(\vect{X}\right)\otimes\vect{N}\left(\vect{X}\right),\label{eq:relation I/I} \\
    \tens{H}_{s}\left(\vect{x}\right) & =\tens{I}\left(\vect{x}\right)-\vect{n}\left(\vect{x}\right)\otimes\vect{n}\left(\vect{x}\right),\label{eq:relation i/I}
\end{align}
where $\tens{I}$ is the identity tensor. We can now define the bulk and the surface
deformation gradient as
\begin{equation}
    \label{eq:deformation_gradients}\tens{F}= \nabla \vect{\chi},\qquad \tens
    {F}_{s}=\tens{F}\,\tens{I}_{s},
\end{equation}
where $\nabla$ denotes the gradient operator using referential coordinates.
Similarly, we introduce a surface determinant operator, indicated with $\det
_{s}$, that accounts for the local area change induced by $\tens{F}_{s}$, see
\cref{appendix:geometry}. In this respect, we set
\begin{equation}
    \label{eq:Js}J_{s}\coloneqq \det_{s}\tens{F}_{s};
\end{equation}
see Appendix \ref{appendix:geometry} for a definition
of the surface determinant and some recalls on differential calculus on material
surfaces.

Since $\tens{I}_{s}$ is a rank-deficient tensor, the surface deformation gradient
$\tens{F}_{s}$ is non-invertible. To overcome this issue, we introduce a
generalized inverse for a general rank-deficient tensor $\tens{A}$ following \cite{holland}.
To this end, we exploit the singular value decomposition
\begin{equation}
    \label{eq:single_value_decomposition}\tens{A}=\tens{V}\,\tens{\Sigma}\,\tens
    {W}^{T},
\end{equation}
where $\tens{\Sigma}$ is a diagonal tensor whose diagonal components
correspond to the singular values of $\tens{A}$, while $\tens{V}$ and $\tens{W}$
represent the tensors whose columns are the left- and right-singular vectors,
respectively. The generalized inverse can be defined as
\begin{equation}
    \label{eq:generalized-inverse}\tens{A}^{-1}=\tens{W}\left[\tens{\Sigma}^{+}
    \right]^{-1}\tens{V}^{T},
\end{equation}
where $\left[\tens{\Sigma}^{+}\right]^{-1}$ is the pseudoinverse of
$\tens{\Sigma}$ obtained by substituting each non-zero entry on the diagonal of
$\tens{\Sigma}$ with its reciprocal value. By construction, we observe that
\citep{Yu_Fu,javili_et_al}
\begin{align*}
    \tens{F}_{s}^{-1}\tens{F}_{s}=\tens{I}_{s}, &  & \tens{F}_{s}\tens{F}_{s}^{-1}=\tens{H}_{s}.
\end{align*}
We are now ready to describe the mechanics of elastic surfaces.

\subsection{Pre-stretched elastic surfaces: balance equations}
\label{subsec:prestretched_surfaces}

\begin{figure}[b!]
    \centering
    \includegraphics[width=0.5\linewidth]{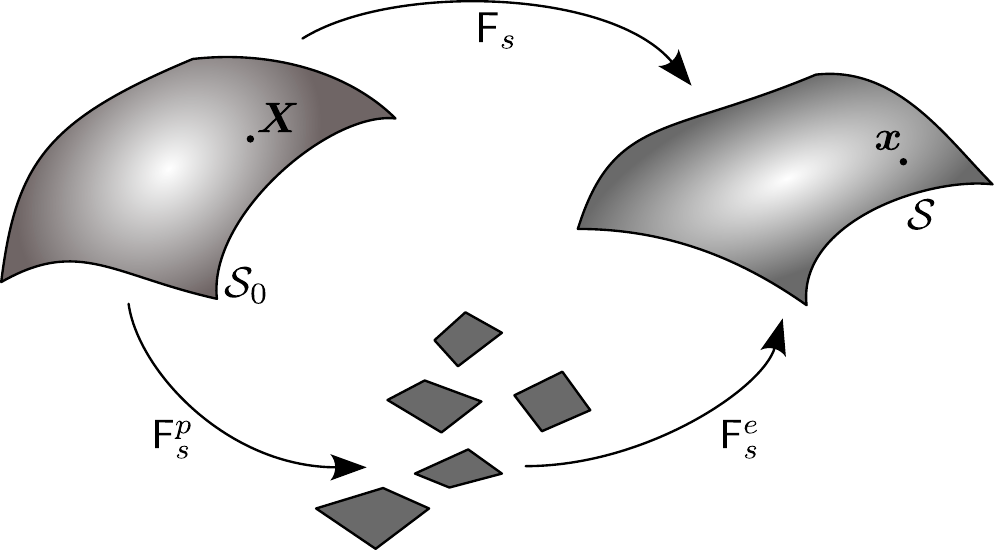}
    \caption{Representation of the multiplicative decomposition of the surface defomation gradient tensor $\tens{F}_s=\tens{F}^e_s\tens{F}^p_s$.}
    \label{fig:decomposition}
\end{figure}

In what follows, we present a model of pre-stretched elastic surfaces surrounding
a three-dimensional continuum. To this end, inspired by the work of \cite{holland},
we can perform a multiplicative decomposition of the surface deformation
gradient in a similar fashion as is done in bulk elasticity
\begin{equation}
    \label{eq:F_s=F_eF_a}\tens{F}_{s}=\tens{F}_{s}^{e}\,\tens{F}_{s}^{p},
\end{equation}
where $\tens{F}_{s}^{p}$ accounts for the elastic pre-stretch, while $\tens{F}_{s}^{e}$ is the elastic distortion from the relaxed state to the current configuration. Specifically, $\tens{F}_s^p$ describes the local distortion of each point on the referential surface to its relaxed state, see \cref{fig:decomposition}.

We assume quasistatic deformations, so that inertia terms can
be neglected. Let $\tens{P}_{s}$ be the first surface Piola-Kirchhoff stress
tensor and $\vect{b}_{0}$ the density of body force per unit referential area.
From the balance of forces we obtain \citep{Gurtin_Murdoch}
\begin{align}
    \label{eq:div P = P N}\nabla_{s}\cdot{\tens{P}}_{s}+{\vect{b}}_{s}=\tens{P}
    \,\vect{N} && \text{on }\mathcal{S}_{0},
\end{align}
where $\tens{P}$ is the (bulk)
Piola-Kirchhoff stress tensor, while $\nabla_{s}\cdot$ denotes the surface divergence operator (see Appendix \ref{appendix:geometry}
for a definition). We remark that \cref{eq:div P = P N} provides the boundary
condition for the classical balance equation of linear momentum of a three dimensional
body
\begin{equation}
    \label{eq:divP + b = 0}\nabla\cdot\tens{P}+\vect{b}= \vect{0}.
\end{equation}

In isothermal conditions, if we assume the existence of a surface strain
energy density per unit reference area $\psi_{s}= \psi_{s}(\tens{F}_{s})$, standard thermodynamic
considerations allow us to write \citep{dehghany2020thermodynamically}
\begin{equation}
    \label{eq:P_s=d_psi_s_d_F_s}\tens{P}_{s}=\frac{\partial \psi_{s}}{\partial
    \tens{F}_{s}}.
\end{equation}
This equation characterizes hyperelastic material surfaces, similarly to
classical hyperelastic materials. In order to construct the strain energy density of the material in the presence of a pre-stretch, a careful treatment is required. 

The boundary layer generated by capillary forces is very small, so we are allowed to consider only stretching effects and neglect bending energy, as in the theoretical framework of \cite{Gurtin_Murdoch}.
It is important to track the variations of the thickness of the material surface in the process.
We assume that the material surface represents the limit of a thin layer of a three-dimensional incompressible material.
Let $H$ and $H_r$ be the (infinitesimal) thicknesses of the material surface in the reference and in the relaxed state, respectively. We recall that $\det_s\tens{F}_s^p$ describes the change in surface area from the reference to the relaxed state, so that
\begin{equation}\label{eq:change_of_variables}
    \mathrm{d}S_r = \det_s \tens{F}_s^p\,\mathrm{d}S,
\end{equation}
where $\mathrm{d}S_r$ and  $\mathrm{d}S$ are the infinitesimal area element in the relaxed and reference state, respectively.
Thus, from the incompressibility of the material we have
\[
H = H_r \det_s \tens{F}_s^p.
\]
We consider the strain energy density $\psi_s^r$ per unit surface area in the relaxed state. We introduce $W_s^r$, defined as 
\[
W_s^r(\tens{F}_s^e) \coloneqq \frac{\psi_s^r(\tens{F}_s^e)}{H_r}.
\]
Here, $W_s^r$ can be regarded as strain energy density per unit volume averaged across the material surface thickness. By \cref{eq:change_of_variables}, introducing $\psi_{s0}\left(\tens{F}_s^e\right) = W_s^r\left(\tens{F}_s^e\right) H$, we get
\[
    \psi_s^r\left(\tens{F}_{s}{\left(\tens{F}_s^p\right)}^{-1}\right)\,\mathrm{d}S_r=\psi_{s0}\left(\tens{F}_{s}{\left(\tens{F}_s^p\right)}^{-1}\right)\,\mathrm{d}S=\psi_s(\mathsf{F}_s)\,\mathrm{d}S.
\]
Therefore, a direct computation shows that
\begin{equation}
    \label{eq:P_s_prestretch}
\tens{P}_{s}=\frac{\partial \psi_{s0}}{\partial
\tens{F}_{s}^{e}}\left(\tens{F}_{s}^{p}\right)^{-T}.
\end{equation}
We can now specialize this framework to a solid cylinder coated by a pre-stretched membrane.

\section{Solid elastic cylinder surrounded by a pre-tensioned elastic
surface}
\label{subsec:model}
Let $\mathscr{B}_{0}\subset\mathbb{R}^{3}$ be the reference
configuration representing the cylinder, with $R_{0}$ denoting its radius.
We introduce the referential cylindrical coordinates $(R,\,\Theta,\,Z)$ and
the corresponding vector basis
$(\vect{E}_{R},\,\vect{E}_{\Theta},\,\vect{E}_{Z})$. We assume that its axial
length is much greater than the radius, so that we can assume
$\mathscr{B}_{0}$ to be infinite in the direction $\vect{E}_{Z}$.

We introduce the current position
$\vect{x}\in \mathscr{B}\subset \mathbb{R}^{3}$ of a point $\vect{X}$, where
$\vect{x}= \vect{\chi}\left(\vect{X}\right)$ and $\mathscr{B}=\vect{\chi}\left(\mathscr{B}
_{0}\right)$ is the current configuration of the cylinder. Moreover, let $(r,\, \theta,\, z)$ be the cylindrical coordinate system in the current configuration.
The corresponding orthonormal basis vectors is given by
$(\vect{e}_{r},\, \vect{e}_{\theta},\, \vect{e}_{z})$.
We denote by $\vect{u}:\mathscr{B}_{0}\rightarrow \mathbb{R}^{3}$ the displacement
field, so that
$\vect{x}\left(\vect{X}\right)=\vect{\chi}\left(\vect{X}\right)=\vect{X}+\vect
{u}\left(\vect{X}\right)$.
The cylinder is free of body forces, so that the balance equations \eqref{eq:div P = P N}
and \eqref{eq:divP + b = 0} become
\begin{subequations}
    \label{sys:eq_conditions}
    \begin{align}
        \nabla\cdot\tens{P}         & =\vect{0}         &  & \text{in }\mathscr{B}_{0},\label{eq:divP=0}    \\
        \nabla_{s}\cdot\tens{P}_{s} & =\tens{P}\vect{N} &  & \text{on }\mathcal{S}_{0}.\label{eq:div_sP_s=PN}
    \end{align}
\end{subequations}
Furthermore, we assume that the system is elongated by a mean stretch
$\lambda$ acting along the $Z$ direction.

In order to proceed with our analysis
we have to make some constitutive assumptions.
The material composing the cylinder volume is assumed to be
incompressible, implying
\begin{equation}
    \label{eq:J=1}J\coloneqq \det\tens{F}= 1.
\end{equation}
We take a neo-Hookean volumetric strain energy density, so that
\begin{equation}
    \label{eq:bulk_energy_density}\psi=\frac{\mu}{2}\left(I_{1}-3\right),
\end{equation}
where $\mu$ is the bulk shear modulus and
$I_{1}=\tr\left(\tens{F}^{T}\tens{F}\right)$.

We assume that the elastic surface is isotropically stretched with
\begin{equation}
    \label{eq:F_a}\tens{F}_{s}^{p}=\lambda_{p}\mathsf{I}_{s},
\end{equation}
where $\lambda_{p}$ is the stretch from the reference to the relaxed state of the body, see \cref{fig:decomposition}. Here, we assume that $\lambda_{p}\in\left(0,1\right
]$ so that the
elastic surface is under tension. The case $\lambda_{p}=1$ corresponds to the absence of surface stress in
the reference configuration.

As a surface strain energy density, we use
\begin{equation}
    \label{eq:surface_energy_density}\psi_{s0}(\tens{F}_s^e)=\frac{\mu_{s}}{2}\bigg(I_{s}^{e}
    -2-2\ln J_{e}\bigg)+\frac{\Lambda_{s}}{2}\left(\frac{1}{2}\left(J_{e}^{2}
    -1\right)-\ln J_{e}\right),
\end{equation}
where $\mu_{s}$ is the surface shear modulus and $\Lambda_{s}$ modulates
surface extensibility (the higher $\Lambda_s$, the more inextensible). The quantity $I_{s}^{e}$ is defined as $I_{s}^{e}=\tr\left
(\left(\tens{F}_{s}^{e}\right)^{T}\tens{F}_{s}^{e}\right)$, while
$J_{e}= \det_s\tens{F}_{s}^{e}$ represents the elastic part of the surface Jacobian $J_{s}$.

Since we are dealing with hyperelastic materials, the bulk
Piola-Kirchhoff stress tensor is given by
\begin{equation}
    \label{eq:P}\tens{P}=\frac{\partial\psi}{\partial\tens{F}}-p\tens{F}^{-T}
    =\mu\tens{F}-p\tens{F}^{-T},
\end{equation}
where $p$, usually called pressure, is a Lagrange multiplier introduced to enforce
the incompressibility constraint.%

\begin{figure}[t!]
    \centering
    \includegraphics[width=0.5\linewidth]{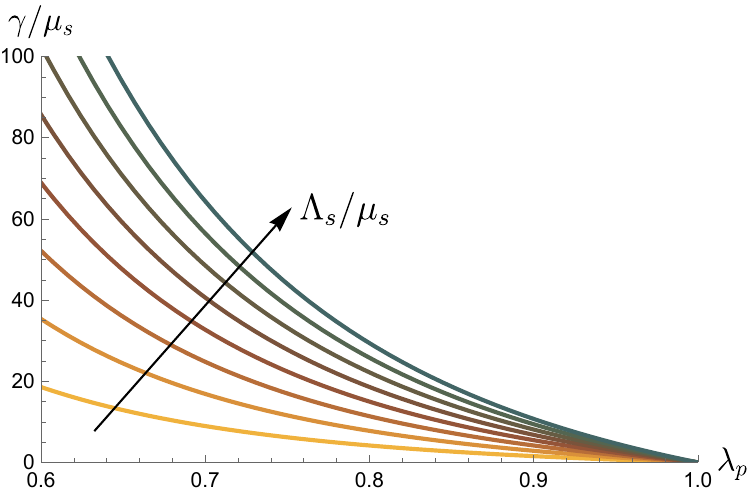}
    \caption{Plot of the surface tension $\gamma$ in the reference configuration, nondimensionalized with respect to the surface shear modulus $\mu_s$, as a function of the pre-stretch parameter $\lambda_p$ for $\Lambda_s/\mu_s=5,\,10,\,\dots,\,40$. The ratio $\Lambda_s/\mu_s$ represents the relative importance of surface extensibility with respect to shear stresses: the higher it is, the more under tension the surface is. The arrow denotes the direction in which $\Lambda_s/\mu_s$ grows.}
    \label{fig:gamma}
\end{figure}

Assuming the multiplicative decomposition \eqref{eq:single_value_decomposition} for
$\tens{F}_{s}$ and using \cref{eq:P_s_prestretch}, we
obtain the following expression for the surface Piola-Kirchhoff stress tensor
\begin{equation}
\label{eq:Ps}
    \tens{P}_{s}= \mu_{s}\left(\tens{F}_{s}\left(\tens{F}_{s}^{p}\right)^{-1}
    \left(\tens{F}_{s}^{p}\right)^{-T}-\tens{F}_{s}^{-T}\right)+\frac{\Lambda_{s}}{2}
    \left(\frac{J_{s}}{J_{s}^{p}}-1\right)\tens{F}_{s}^{-T}.
\end{equation}

\begin{remark}
\label{rem:st_lambdap}
    In the undeformed reference configuration, i.e. when $\tens{F}=\tens{I}$, the surface stress $\tens{P}_s$ corresponds to an isotropic surface stress, as in fluids, so that
    \[
    \tens{P}_s = \gamma \tens{I}_s.
    \]
    We call $\gamma$ \emph{initial surface tension}.
    It is a function of $\lambda_p$ and the material parameters $\mu_s$ and $\Lambda_s$, specifically
    \begin{equation}
        \label{eq:gamma}
        \gamma = \frac{1-\lambda_p^4}{2\lambda_p^4}\Lambda_s+\frac{1-\lambda_p^2}{\lambda_p^2} \mu_s.
    \end{equation}
    In particular, the surface is under tension for $0<\lambda_p<1$, see \cref{fig:gamma}.
    For small deformations from the reference configuration, the stress $\tens{P}_s$ obeys to the Shuttleworth equation, as discussed by \cite{Yu_Fu}, with the initial surface tension $\gamma$ given by \cref{eq:gamma}.
\end{remark}

In the next section, we show that the cylinder can undergo a beading instability when subjected to a homogeneous uniaxial
stretch.

\section{Stability analysis}
\label{sec:analysis_of_the_model} In this section we conduct a linear
stability analysis of the cylindrical configurations with respect to axisymmetric perturbations with a wavenumber $k$. In what follows, we show that the body always admits a cylindrical solution, and we perform a stability analysis. Readers mainly interested in results are suggested to go directly to \cref{subsec:results_lsa}.

\subsection{Cylindrical solution}
\label{subsec:cylindrical_solutions} Cylindrical solutions
representing homogeneous uniaxial extensions are represented by the following class of deformations
\begin{equation}
    \label{eq:hom_def_incompr}r = \frac{R}{\sqrt{\lambda}}, \qquad z = \lambda
    Z,
\end{equation}
which satisfies the incompressibility constraint (\ref{eq:J=1}). Indeed, the deformation gradient associated to \cref{eq:hom_def_incompr} is given by
\[
    \tens{F}=\lambda^{-\frac{1}{2}}\left(\vect{e}_{r}\otimes\vect{E}_{R}+\vect
    {e}_{\theta}\otimes\vect{E}_{\Theta}\right)+\lambda\vect{e}_{z}\otimes\vect
    {E}_{Z}.
\]
The corresponding surface deformation gradient can be obtained through
\cref{eq:deformation_gradients}, so that
\[
    \tens{F}_{s}=\lambda^{-\frac{1}{2}}\vect{e}_{\theta}\otimes\vect{E}_{\Theta}
    +\lambda\vect{e}_{z}\otimes\vect{E}_{Z}.
\]
From \cref{eq:P}, we compute the bulk Piola-Kirchhoff stress tensor
\begin{equation}
\label{eq:Pcyl}
    \tens{P}= \left(\frac{\mu }{\sqrt{\lambda }}-\sqrt{\lambda }p\right)(\tens
    {I}-\vect{e}_{z}\otimes\vect{E}_{Z})+\left(\lambda \mu -\frac{p}{\lambda }
    \right)\vect{e}_{z}\otimes\vect{E}_{Z}
\end{equation}
and its surface counterpart through \cref{eq:Ps}
\begin{equation}
\label{eq:Pcyl_s}
    \tens{P}_{s}= \left(\frac{1}{2}\left(\frac{\lambda^{3/2}}{\lambda_{p}^{4}}
    -\sqrt{\lambda }\right) \Lambda_{s}+\left(\frac{1}{\sqrt{\lambda }\lambda_{p}^{2}}
    -\sqrt{\lambda }\right) \mu_{s}\right)\vect{e}_{\theta}\otimes\vect{E}_{\Theta}
    + \left( \frac{1}{2}\left(\frac{1}{\lambda_{p}^{4}}-\frac{1}{\lambda }\right
    ) \Lambda_{s}+\left(\frac{\lambda }{\lambda_{p}^{2}}-\frac{1}{\lambda }\right
    ) \mu_{s}\right)\vect{e}_{z}\otimes\vect{E}_{Z}.
\end{equation}
Finally, we can find the expression of the pressure field $p$ by enforcing the boundary condition \cref{eq:div_sP_s=PN}. Using \cref{eq:Pcyl,eq:Pcyl_s} we get
\begin{equation*}
    \label{eq:P_RR=-Ps_TT/R_0}P_{RR}=-\frac{P_{s \, \Theta\Theta}}{R_{0}}.
\end{equation*}
By solving this equation with respect to $p$ we obtain
\begin{equation}
    \label{eq:p_hom_def}p=\frac{2 \lambda_{p}^{2}\left(\mu R_{0}\lambda_{p}^{2}+\mu_{s}\right)-\lambda
    \lambda_{p}^{4}\left(\Lambda_{s}+2 \mu_{s}\right)+\lambda^{2}\Lambda_{s}}{2
    \lambda R_{0}\lambda_{p}^{4}}.
\end{equation}
We can now investigate possible bifurcations of the cylindrical
configuration that can eventually lead to a beading instability.

\subsection{Incremental relations}
We make use of the theory of incremental deformations to analyse the linear stability of the cylindrical configuration \citep{ogden1997non}.
We introduce the incremental displacement and pressure fields, denoted by $\delta\vect{u}:\mathscr{B}\rightarrow
\mathbb{R}^{3}$ and $\delta p:\mathscr{B}\rightarrow
\mathbb{R}$, respectively. We set $\tens{\Gamma}\coloneqq\grad \delta\vect{u}$.

Similarly, let $\tens{\Gamma}_{s}$ be the surface gradient of the
incremental displacement, so that $\tens{\Gamma}_{s}=\tens{\Gamma}\,\tens{H}_{s}$.
The bulk and surface incremental Piola-Kirchhoff stress tensors are given by \citep{Yu_Fu}
\begin{align}\label{dP_s_C0}
        \begin{aligned}
            \delta\tens{P}     & =\bm{\mathcal{A}}_{0}:\tens{\Gamma}+p\,\tens{\Gamma}^{T}-\delta p\,\tens{I}, \\
        \delta\tens{P}_{s} & =\bm{\mathcal{C}}_{0}:\tens{\Gamma}_{s}, 
        \end{aligned}
        &&
        \text{so that}
        &&
        \begin{aligned}
            \delta P_{ij}     & ={\mathcal{A}}_{0\,ijhk}{\Gamma_{hk}}+p\,{\Gamma_{ji}}-\delta p\,{I_{ij}},\\
        \delta{P}_{s\,ij} & ={\mathcal{C}}_{0\,ijhk}{\Gamma}_{s\,hk}. 
        \end{aligned}
\end{align}
where $\bm{\mathcal{A}}_{0}$ and $\bm{\mathcal{C}}_{0}$ are the fourth-order
tensors of the bulk and surface instantaneous elastic moduli, respectively.
Their components are given by
\begin{subequations}
    \label{sys:elstic_moduli}
    \begin{align}
        \mathcal{A}_{0\,ijhk} & ={F}_{jm}{F}_{kn}\frac{\partial^{2}\psi}{\partial F_{im}\partial F_{hn}},\label{eq:bulk_elastic_modulus_A0}                                  \\
        \mathcal{C}_{0\,ijhk} & =J_{s}^{-1}{F}_{jm}^{s}{F}_{kn}^{s}\frac{\partial^{2}\psi_{s}}{\partial F_{im}^{s}\partial F_{hn}^{s}},\label{eq:surface_elastic_modulus_C0}
    \end{align}
\end{subequations}
where we assume summation over repeated indices. The incremental counterpart
of \cref{eq:divP=0,eq:div_sP_s=PN} and \cref{eq:J=1} are given by
\begin{subequations}
    \label{sys:incremental_eq_conditions}
    \begin{align}
        \diver \delta\tens{P}        & =\vect{0}                 &  & \text{in }\mathscr{B},\label{eq:incr_divP=0}      \\
        \tr\tens{\Gamma}             & =0                 &  & \text{in }\mathscr{B},\label{eq:tr_Gamma=0}       \\
        \diver_{s}\delta\tens{P}_{s} & =\delta\tens{P}\,\bm{n} &  & \text{on }\mathcal{S}.\label{eq:incr_div_sP_s=PN}
    \end{align}
\end{subequations}

If the material is isotropic, a convenient way of computing the components of the
tensors $\bm{\mathcal{A}}_{0}$ and $\bm{\mathcal{C}}_{0}$ is to rely on the principal
stretches, indicated in this case by
$\lambda_{r}= \lambda_{\theta}= \lambda^{-1/2}$ and $\lambda_{z}=\lambda$. Indeed,
by using cylindrical coordinates with $i,j\in\{r,\theta,z\}$ and
$\alpha,\beta\in\{\theta,z\}$ we get \citep{ogden1997non}
\begin{subequations}
    \label{eq:A0_isotropic}
    \begin{align}
        \mathcal{A}_{0\,iijj} & =\lambda_{i}\lambda_{j}\psi_{,ij},                                                                                                     \\
        \mathcal{A}_{0\,jiji} & =\frac{\lambda_{i}^{2}}{\lambda_{i}^{2}-\lambda_{j}^{2}}\left(\lambda_{i}\psi_{,i}-\lambda_{j}\psi_{,j}\right) \quad \text{if }i\neq j, \\
        \mathcal{A}_{0\,ijji} & =\mathcal{A}_{0\,jiij}=\mathcal{A}_{0\,jiji}-\lambda_{i}\psi_{,i} \quad\text{if }i\neq j,
    \end{align}
\end{subequations}
while, for the surface elastic moduli we obtain \citep{Yu_Fu,chadwick_ogden}
\begin{equation*}
    \label{eq:C0_isotropic}
    \begin{aligned}
        J_{s}\mathcal{C}_{0\,\alpha\alpha\beta\beta} & =\lambda_{\alpha}\lambda_{\beta}\psi_{s,\alpha\beta},                                                                                                                               \\
        J_{s}\mathcal{C}_{0\,\beta\alpha\beta\alpha} & =\frac{\lambda_\alpha^2}{\lambda_\alpha^2-\lambda_\beta^2}\left(\lambda_{\alpha}\psi_{s, \alpha}-\lambda_{\beta}\psi_{s, \beta}\right) \quad \text{if }\alpha\neq\beta,            \\
        J_{s}\mathcal{C}_{0\,\beta\alpha\alpha\beta} & =\frac{\lambda_\alpha\lambda_\beta}{\lambda_\alpha^2-\lambda_\beta^2}\left(\lambda_{\beta}\psi_{s, \alpha}-\lambda_{\alpha}\psi_{s, \beta}\right) \quad \text{if }\alpha\neq\beta, \\
        J_{s}\mathcal{C}_{0\,r \alpha r\alpha}      & =\lambda_{\alpha}\psi_{s, \alpha}.
    \end{aligned}
\end{equation*}

\subsection{Axisymmetric solutions of the incremental problem and linear
stability analysis}
We can now proceed with the solution of the incremental
problem \eqref{sys:incremental_eq_conditions} and the construction of a bifurcation criterion. 

While an axially compressed cylinder can undergo a non-axisymmetric instability, similarly to the classical Euler buckling problem \citep{Goriely_2008}, in the presence of surface tension and axial traction the buckling mode is axisymmetric \citep{PhysRevLett.105.214301, fu2021necking}.
Therefore, in the following we focus on axisymmetric
perturbations to the base solution, i.e. we assume that the incremental
displacement $\delta\vect{u}$ and the incremental pressure $\delta p$ have the following structure:
\begin{equation}
    \label{eq:du_dp}\delta\vect{u}=u\left(r,\,z\right)\vect{e}_{r}+w\left(r,\,z\right)\vect{e}_{z},\qquad \delta p = \delta p(r,\,z).
\end{equation}

By exploiting a matrix representation of second order tensors through the
cylindrical basis, we can write the gradient of the incremental displacement
as
\begin{equation}
    \label{eq:Gamma_surface}\tens{\Gamma}=\grad\delta\vect{u}=
    \begin{pmatrix}
        \frac{\partial u}{\partial r} & 0           & \frac{\partial u}{\partial z} \\
        0                             & \frac{u}{R} & 0                             \\
        \frac{\partial w}{\partial r} & 0           & \frac{\partial w}{\partial z}
    \end{pmatrix}.
\end{equation}
The surface counterpart of $\tens{\Gamma}$ can be obtained through \cref{eq:relation i/I}, so that
\begin{equation*}
    \tens{\Gamma}_s=\tens{\Gamma}\,\tens{H}_{s}=
    \begin{pmatrix}
        0 & 0           & \frac{\partial u}{\partial Z} \\
        0 & \frac{u}{R} & 0                             \\
        0 & 0           & \frac{\partial w}{\partial Z}
    \end{pmatrix}.
\end{equation*}

In order to proceed with the analysis, we assume the following separation of variables
\begin{subequations}
    \label{sys:variable_separation}
    \begin{align}
        u\left(r,z\right)        & =U(r)\sin{\left(kz\right)}\label{eq:u=U_sin(kz)},  \\
        w\left(r,z\right)        & =W(r)\cos{\left(kz\right)}\label{eq:w=W_sin(kz)},  \\
        \delta p\left(r,z\right) & =Q(r)\sin{\left(kz\right)}\label{eq:dp=Q_sin(kz)},
    \end{align}
\end{subequations}
where $k$ is the wavenumber of the perturbation. Hence, we can obtain $W(r)$ and $Q(r)$ as a function of $U(r)$ and its derivatives from
\cref{eq:tr_Gamma=0} and the expression of $\delta P_{rr}$ in \cref{dP_s_C0}, respectively, so that
\begin{align*}
    W\left(r\right)&=\frac{r U'(r)+U(r)}{k r},\\
    Q\left(r\right)&=\frac{\mu  \left(r \left(r \left(r U^{(3)}(r)+2 U''(r)\right)-\left(k^2 \lambda ^3 r^2+1\right) U'(r)\right)+U(r) \left(1-k^2 \lambda ^3
   r^2\right)\right)}{k^2 \lambda  r^3}.
\end{align*}
From \cref{eq:incr_divP=0}, we finally obtain a fourth order ordinary differential equation for $U(r)$:
\begin{equation}
    \label{eq:diff_eq_U}
    \begin{aligned}
        \frac{r \left(\left(k^2 \left(\lambda ^3+1\right) r^2-3\right) U'(r)+r \left(\left(k^2 \left(\lambda ^3+1\right) r^2+3\right) U''(r)-r \left(r U^{(4)}(r)+2 U'''(r)\right)\right)\right)}{k^2 \lambda r^4}+ &     \\
        -\frac{U(r) \left(k^4 \lambda ^3 r^4+k^2 \left(\lambda ^3+1\right) r^2-3\right)}{k^2 \lambda r^4}                                                                                                              & =0.
    \end{aligned}
\end{equation}

The general solution of \cref{eq:diff_eq_U} for $U(r)$, ensuring the continuity of the body along the cylinder's axis, is given by a linear combination of two independent functions, $U_{1}(r)$
and $U_{2}(r)$ \citep{BIGONI20015117}, i.e. 
\begin{equation}
\label{eq:U(r)}
    U(r)=c_{1}U_{1}(r)+c_{2}U_{2}(r),
\end{equation}
where $c_{1}$ and $c_{2}$ are arbitrary constants. In particular, when
$\lambda\neq 1$,
\begin{equation}
    \label{eq:U(R)_lambda!=1}U_1(r)=J_{1}(k r q_{1}
    )\qquad U_2(r)=J_{1}(k r q_{2}),
\end{equation}
where $J_{m}$ is the modified Bessel function of the first kind of order $m$,
while $q_{1}$ and $q_{2}$ are two coefficients given by
\begin{equation}
    \label{eq:q1_q2}q_{1,\,2}^{2}=\frac{\lambda^{3}+1\pm(\lambda^{3}-1)}{2}
    .
\end{equation}
Instead, if $\lambda=1$, we get \citep{BIGONI20015117}
\begin{equation}
    \label{eq:U(R)_lambda=1}U_1(r)=J_{1}(k r),\qquad U_2(r)=r
    J_{0}(k r).
\end{equation}

We are now left with the imposition of the boundary
condition \cref{eq:incr_div_sP_s=PN}. Given the solution \cref{eq:U(r)}, such a boundary condition reduces to a linear system whose unknowns are $\left(c_{1},\,c_{2}
\right)=\vect{c}$, i.e.
\begin{equation*}
    \tens{M}\vect{c}=\vect{0}.
\end{equation*}
Here, $\tens{M}$ is a $2\times2$ matrix whose elements are reported in \cref{appendix:matrix}.

Non-trivial solutions of the incremental problem exist when the
matrix $\tens{M}$ is singular, that is
\begin{equation}\label{eq:phi=det_M=0}
    \varphi\left(k,\mu,\lambda,\mu_{s},\Lambda_{s},\lambda
    _{p},R_{0}\right)\coloneqq \det\tens{M}=0.
\end{equation}
\Cref{eq:phi=det_M=0} is highly nonlinear and it is not possible
to find analytic expressions for the roots. Thus, we rely on numerical computations
to find its solutions, using a Newton algorithm implemented
with the software \texttt{Mathematica 13.3} (Wolfram Research, Champaign, IL, USA).

In the following section we perform a non-dimensionalization of \cref{eq:phi=det_M=0} and present and discuss the results of the linear stability analysis.

\subsection{Results of the linear stability analysis}\label{subsec:results_lsa}

\begin{figure}[b!]
    \centering
    \subfloat[$\lambda=1$.\label{fig:k_mu_lambda_1}]{ \includegraphics[width=0.45\textwidth]{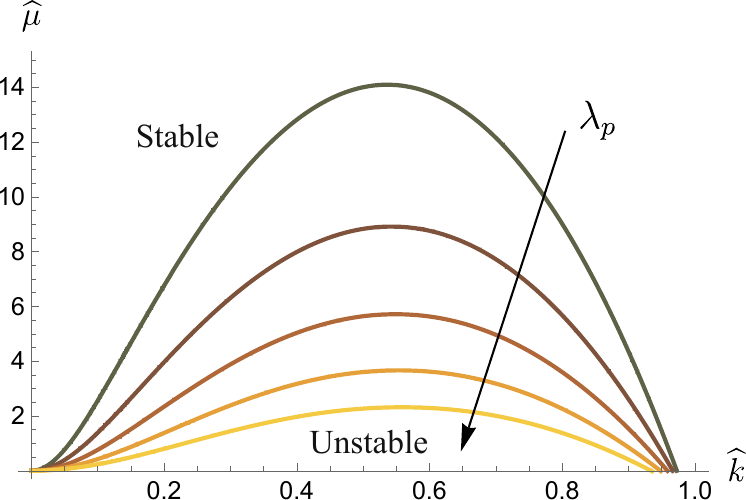} }
    \quad \subfloat[$\lambda=1.4$.\label{fig:k_mu_lambda_not_1}]{ \includegraphics[width=0.45\textwidth]{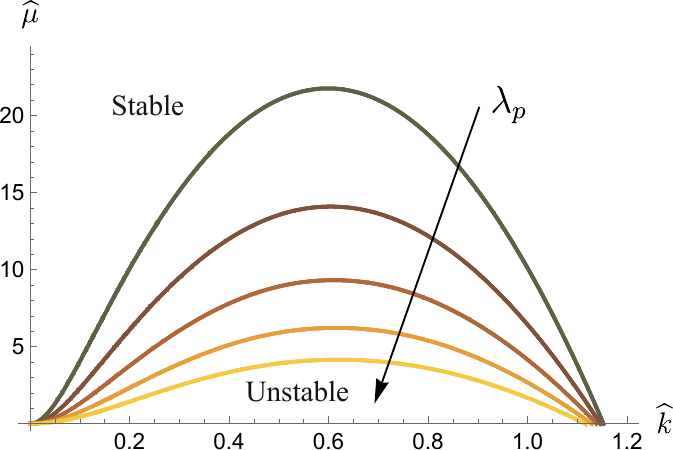} }
    \caption[]{Plot of the dimensionless shear modulus $\widehat{\mu}$ versus the dimensionless wave
    number $\widehat{k}$ for unitary (\cref{fig:k_mu_lambda_1}) and nonunitary (\cref{fig:k_mu_lambda_not_1}) axial strain, respectively. Here, the dimensionless surface extensibility $\widehat{\Lambda}
    _{s}$ is set to 40 and the pre-stretch parameter $\lambda_{p}=0.4,\,0.5,\,0.6,\,0.7,\,0.8$. The arrow
    denotes the direction of growth of $\lambda_p$.}
    \label{fig:linear_stability_analysis_k_mu}
\end{figure}

First, we identify the non-dimensional quantities that govern the behaviour of the system. In particular, we choose
$R_{0}$ and $\mu_{s}$ as characteristic length and stiffness of the system, respectively.
This choice allows us to identify the following dimensionless quantities
\begin{align*}
    \widehat{\mu}=\frac{\mu R_{0}}{\mu_{s}}, &  & \widehat{\Lambda}_{s}=\frac{\Lambda_{s}}{\mu_{s}}, &  & \widehat{k}=k R_{0},
\end{align*}
representing the dimensionless shear modulus, the dimensionless surface extensibility, and the dimensionless wavenumber, respectively. We observe that $\mu_s/\mu$ can be interpreted as the counterpart of the classical elasto-capillary length of the system \citep{style2017elastocapillarity}, while $\widehat{\Lambda}_s$ measures the extensibility of the free surface.

\begin{figure}[t!]
    \centering
    \subfloat[$\lambda=1$.\label{fig:k_mu_la_cr_lambda_1}]{ \includegraphics[width=0.45\textwidth]{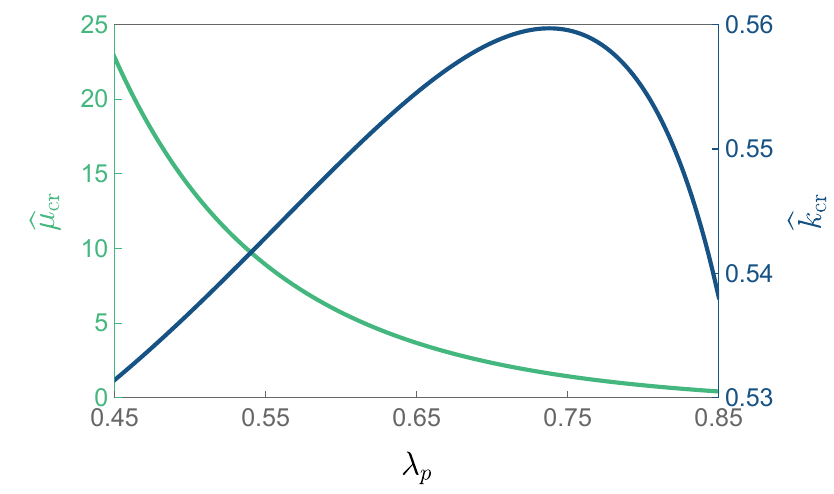} }
    \quad \subfloat[$\lambda=1.4$.\label{fig:k_mu_la_cr_lambda_not_1}]{ \includegraphics[width=0.45\textwidth]{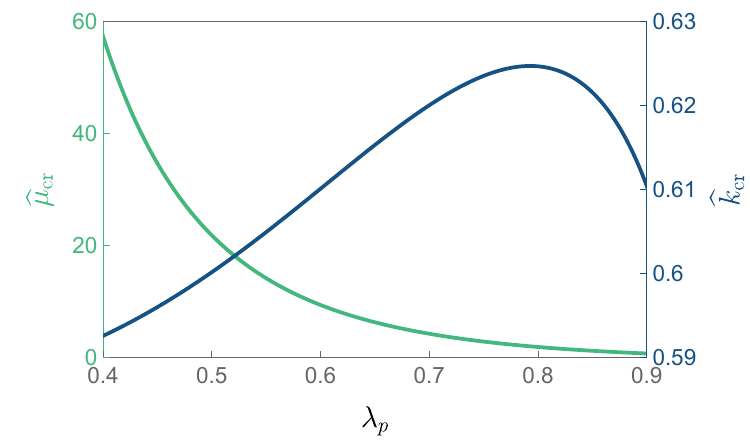} }
    \caption[]{Plot of the critical bifurcation thresholds of the dimensionless shear modulus $\widehat{\mu}_\text{cr}$ (turquoise) and of the dimensionless wavenumber
    $\widehat{k}_\text{cr}$ (blue) against the pre-stretch parameter $\lambda_{p}$ for unitary (\cref{fig:k_mu_la_cr_lambda_1}) and nonunitary (\cref{fig:k_mu_la_cr_lambda_not_1}) axial strain, respectively. Here, the dimensionless surface extensbility $\widehat
    {\Lambda}_{s}$ is set to 40.}
    \label{fig:linear_stability_analysis_k_mu_la_cr}
\end{figure}

\begin{figure}[t!]
    \centering
    \subfloat[$\lambda=1$.\label{fig:k_mu_Ls_lambda_1}]{ \includegraphics[width=0.45\textwidth]{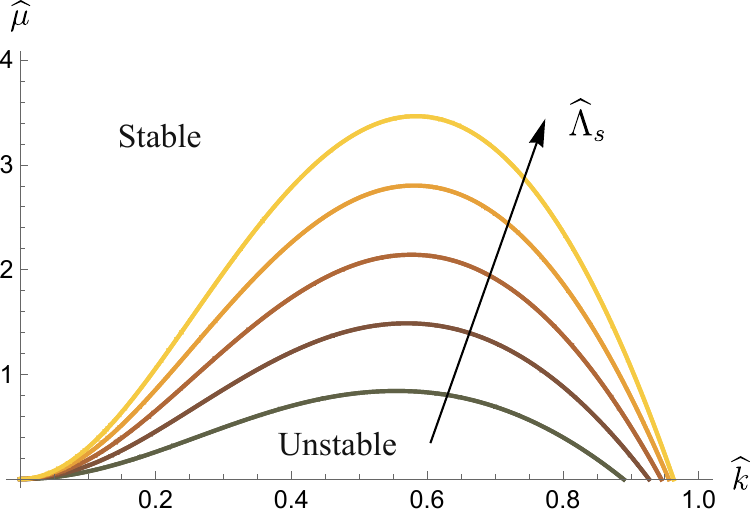} }
    \quad \subfloat[$\lambda=1.4$.\label{fig:k_mu_Ls_lambda_not_1}]{ \includegraphics[width=0.45\textwidth]{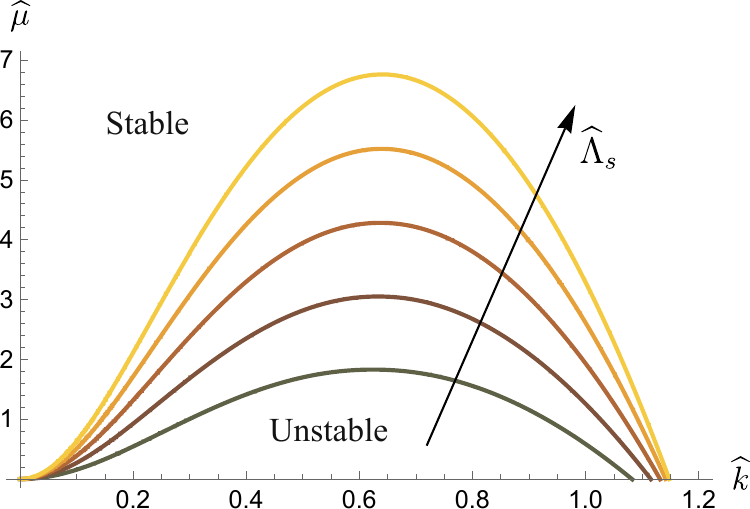} }
    \caption[]{Plot of the dimensionless shear modulus $\widehat{\mu}$ versus the dimensionless wave
    number $\widehat{k}$ for unitary  (\cref{fig:k_mu_Ls_lambda_1}) and nonunitary (\cref{fig:k_mu_Ls_lambda_not_1}) axial strain, respectively. Here, the pre-stretch parameter $\lambda_{p}$ is set to 0.8 and the dimensionless surface extensibility $\widehat{\Lambda}_{s}=40,\,60,\,80,\,100,\,120$. The arrow
    denotes the direction of growth of $\widehat{\Lambda}_{s}$.}
    \label{fig:linear_stability_analysis_k_mu_Ls}
\end{figure}

\begin{figure}[t!]
    \centering
    \subfloat[$\lambda=1$.\label{fig:k_mu_Ls_cr_lambda_1}]{ \includegraphics[width=0.45\textwidth]{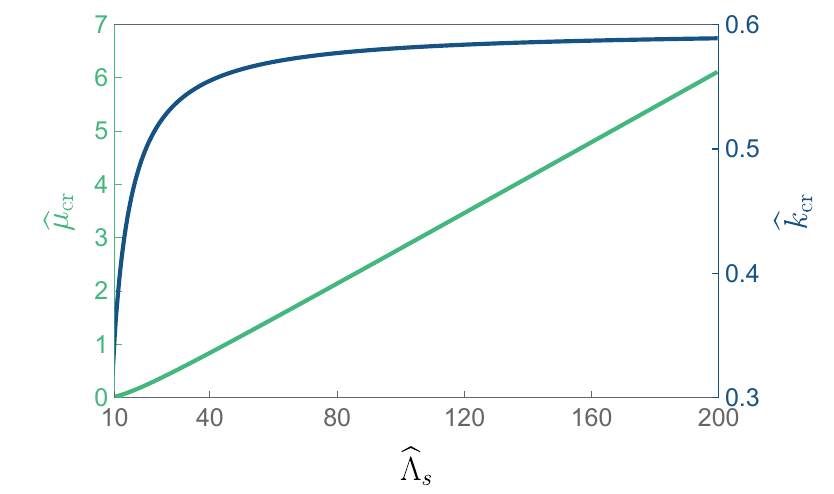} }
    \quad \subfloat[$\lambda=1.4$.\label{fig:k_mu_Ls_cr_lambda_not_1}]{ \includegraphics[width=0.45\textwidth]{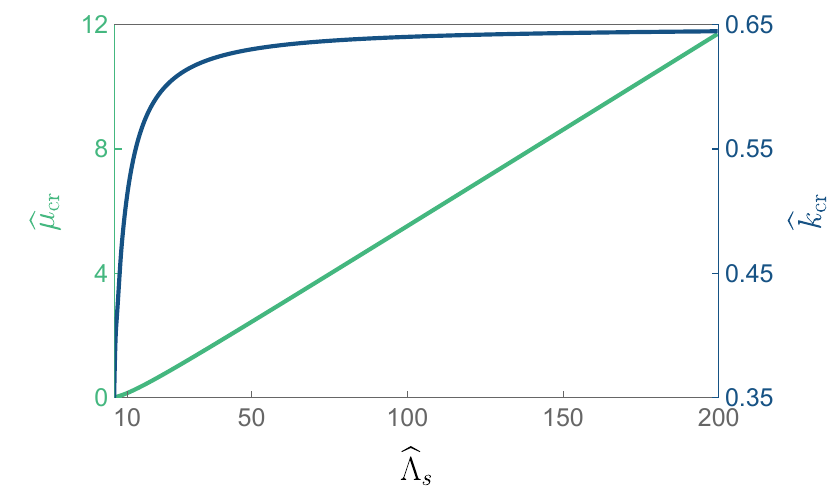} }
    \caption[]{Plot of the critical bifurcation thresholds of the dimensionless shear modulus $\widehat{\mu}_\text{cr}$ (turquoise) and of the dimensionless wavenumber
    $\widehat{k}_\text{cr}$ (blue) against the dimensionless surface extensibility $\widehat{\Lambda}_{s}$ for unitary (\cref{fig:k_mu_Ls_cr_lambda_1}) and nonunitary (\cref{fig:k_mu_Ls_cr_lambda_not_1}) axial strain, respectively. Here, the pre-stretch parameter $\lambda_{p}$ is set to 0.8.}
    \label{fig:linear_stability_analysis_k_mu_Ls_cr}
\end{figure}

In the following, we refer to the first mode that becomes unstable as the critical buckling mode. The corresponding dimensionless critical wavenumber is denoted by $\widehat{k}_\text{cr}$.

We first explore the stability of the cylindrical configuration with respect to control parameter
$\widehat{\mu}$, analysing the effect of surface pre-stretch.
In all the studied cases, as we decrease $\widehat{\mu}$, the critical mode has a non-zero wavenumber, see \cref{fig:linear_stability_analysis_k_mu}. The marginal stability curves in \cref{fig:linear_stability_analysis_k_mu} show a similar trend as we change the parameters. Specifically, we observe a decrease in the critical thresholds for $\widehat{\mu}$ as $\lambda_p$ approaches $1$. Moreover, axial stretching appears to stabilize the cylinder: a bifurcation occurs at larger values of $\widehat{\mu}$ when $\lambda=1.4$ with respect to the case with $\lambda=1$, as shown in \cref{fig:linear_stability_analysis_k_mu}.
In \cref{fig:linear_stability_analysis_k_mu_la_cr} we report the trend of the critical wavenumber and of the critical shear modulus. Our results show that $\widehat{\mu}_\text{cr}$ monotonically decreases as $\lambda_p$ increases.

We now analyse the effect of surface extensibility. The results are reported in
\cref{fig:linear_stability_analysis_k_mu_Ls}. %
As the surface becomes less extensible, i.e. as $\widehat{\Lambda}_s$ grows, also $\widehat{\mu}_\text{cr}$ monotonically grows. From \cref{fig:linear_stability_analysis_k_mu_Ls_cr} we
can notice that this trend is linear for both $\lambda=1$ and $\lambda=1.4$.
In particular, the axial strain has a stabilizing effect on the cylinder, increasing $\widehat{\mu}
_\text{cr}$ as the axial stretch grows. Furthermore, we observe that the critical wavenumber rapidly increases
for small $\widehat{\Lambda}_{s}$, and saturates to a constant value when the surface is nearly inextensible. From a physical standpoint, these results suggest that the periodic beading pattern can be triggered more easily when the surface is stiffer and nearly inextensible. We also remark that, as before, the critical wavenumber predicted by the linear stability analysis is finite while the system remains stable for longer wavelengths ($\widehat{k}\rightarrow 0$). Thus, periodic patterning appears to be favourable compared to isolated bulging, in agreement with the observations in the experiments \citep{PhysRevLett.105.214301}.

\begin{figure}[t!]
    \centering
    \subfloat[$\lambda=1$.\label{fig:k_la_lambda_1}]{ \includegraphics[width=0.45\textwidth]{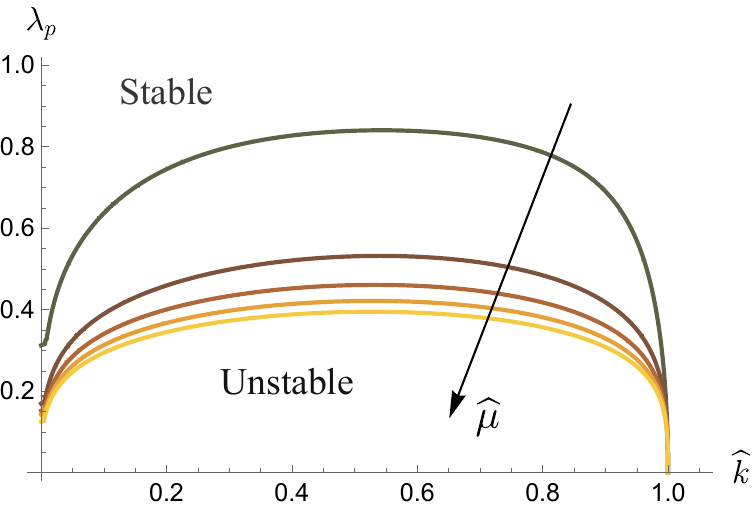} }
    \quad \subfloat[$\lambda=1.4$.\label{fig:k_la_lambda_not_1}]{ \includegraphics[width=0.45\textwidth]{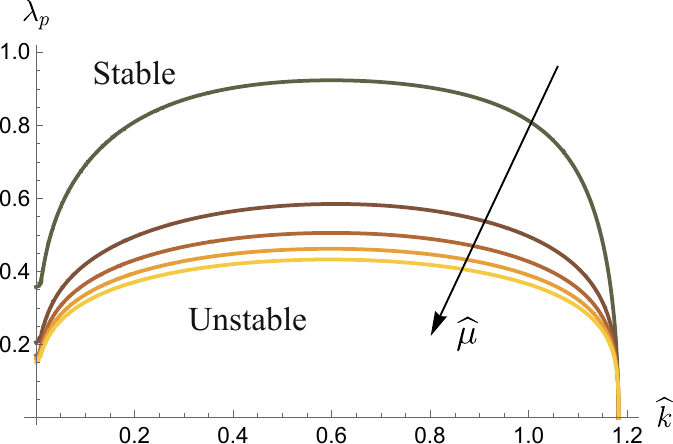} }
    \caption[]{Plot of the pre-stretch parameter $\lambda_{p}$ versus
    the dimensionless wavenumber $\widehat{k}$ for unitary (\cref{fig:k_la_lambda_1})
    and nonunitary (\cref{fig:k_la_lambda_not_1}) axail strain, respectively. Here, dimensionless surface extensibility $\widehat{\Lambda}
    _{s}$ is set to 40 and the dimensionless shear modulus $\widehat{\mu}=0.5,\,10.5,\,20.5,\,30.5,\,40.5$. The arrow
    denotes the direction of growth of $\widehat{\mu}$.}
    \label{fig:linear_stability_analysis_k_la}
\end{figure}

\begin{figure}[t!]
    \centering
    \subfloat[$\lambda=1$.\label{fig:k_la_mu_cr_lambda_1}]{ \includegraphics[width=0.45\textwidth]{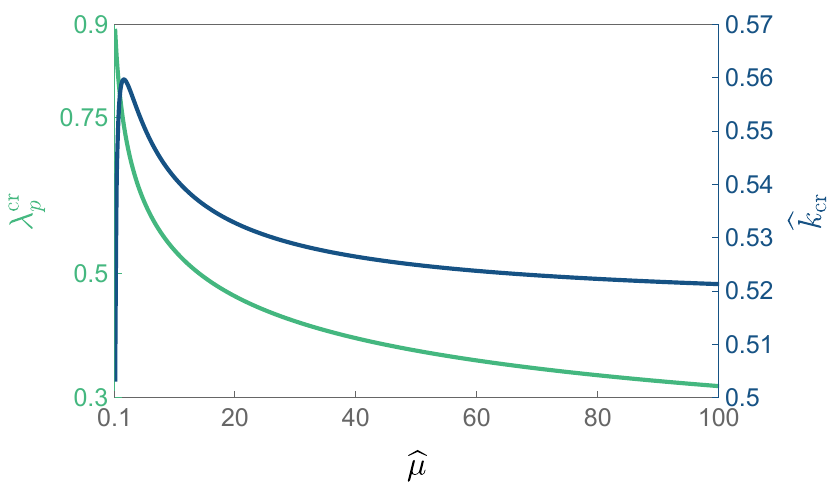} }
    \quad \subfloat[$\lambda=1.4$.\label{fig:k_la_mu_cr_lambda_not_1}]{ \includegraphics[width=0.45\textwidth]{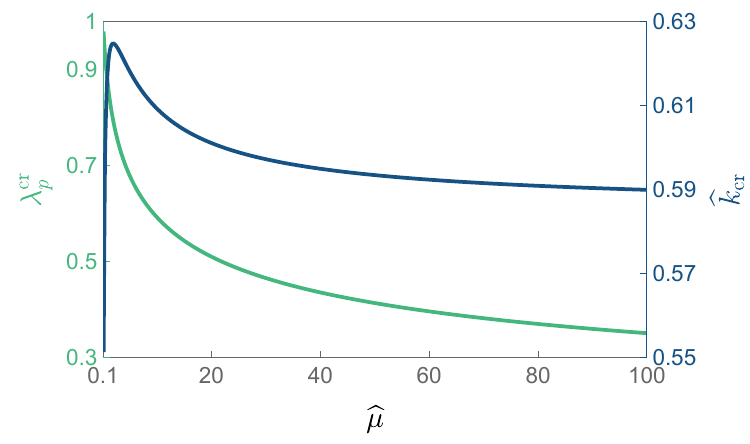} }
    \caption[]{Plot of the critical bifurcation thresholds of the pre-stretch parameter $\lambda_{p}^{cr}$ (turquoise) and of the dimensionless wavenumber $\widehat{k}_\text{cr}$
    (blue) against the dimensionless shear modulus $\widehat{\mu}$ for unitary (\cref{fig:k_la_mu_cr_lambda_1})
    and nonunitary (\cref{fig:k_la_mu_cr_lambda_not_1}) axial strain, respectively. Here, the dimensionless surface extensibility $\widehat
    {\Lambda}_{s}$ is set to 40.}
    \label{fig:linear_stability_analysis_k_la_mu_cr}
\end{figure}

We also explore the stability of cylindrical configurations by modulating the surface tension through $\lambda_{p}$. The results are shown in \cref{fig:linear_stability_analysis_k_la}.
We find again a non-zero critical wavenumber, accordingly with the beading phenomenon. From \cref{fig:linear_stability_analysis_k_la_mu_cr} we can also notice that when $\widehat{\mu}$ decreases, the critical threshold for $\lambda_p$ decreases as well, while the wavenumber sharply increases for small values of $\widehat{\mu}$ and saturates at a constant value when the non-dimensional shear modulus is
sufficiently large.

\begin{figure}[t!]
    \centering
    \includegraphics[width=0.45\textwidth]{
        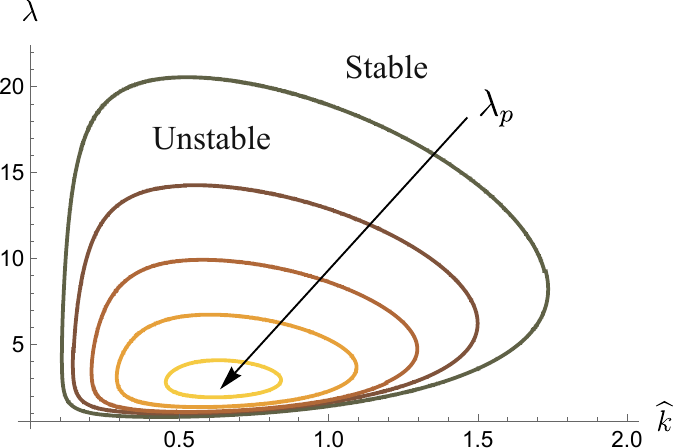
    }
    \caption[]{Plot of the axial stretch $\lambda$ versus the dimensionless wavenumber $k$. The arrow denotes the direction of growth of the pre-stretch parameter $\lambda_p$ with uniform steps of amplitude 0.05, from 0.5 to 0.7.}
    \label{fig:linear_stability_analysis_k_lambda_lambda_p}
\end{figure}

\begin{figure}[t!]
    \centering
    \includegraphics[width=0.45\textwidth]{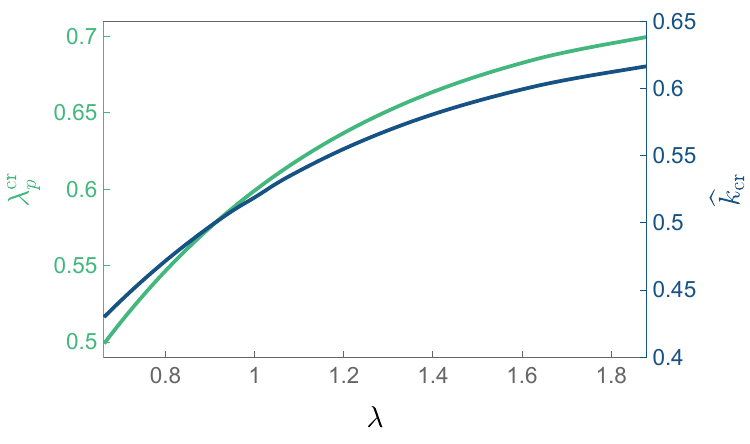}
    \caption[]{Plot of the critical bifurcation thresholds of the pre-stretch parameter $\lambda_{p}^{cr}$ (turquoise) and of the dimensionless wavenumber $\widehat{k}_\text{cr}$
    (blue) against the axial strain $\lambda$. Here, we set the dimensionless surface extensibility $\widehat{\Lambda}_s$ to 10 and the dimensionless shear modulus $\widehat{\mu}$ to 0.8.}
    \label{fig:linear_stability_analysis_k_lambda_lambda_p_cr}
\end{figure}

Finally, we analyse the stability of the cylindrical configuration with respect to the axial strain $\lambda$. As shown in \cref{fig:linear_stability_analysis_k_lambda_lambda_p}, we find a finite positive critical wavenumber, consistently with all other cases explored in this section. Interestingly, the marginal stability curves form closed loops, suggesting that while buckling occurs initially, the system may return to the unbuckled state if $\lambda$ becomes sufficiently large. This aspect will be investigated in the following through finite element simulations. Moreover, from \cref{fig:linear_stability_analysis_k_lambda_lambda_p_cr}, we notice that both the critical thresholds for $\lambda_p$ and for the dimensionless wavenumber increase sublinearly as $\lambda$ is incremented. We observe that the critical wavenumber $\widehat{k}_\text{cr}$ is always between $0.5$ and $0.7$ in all the cases examined in this section. In the next section, we characterize the post-buckling behaviour of the critical mode in the fully nonlinear regime.


\section{Numerical simulations}
\label{sec:numerical_simulations} 
In this section, we detail the numerical methods and present the results of the simulations for the nonlinear boundary value problem \eqref{eq:divP=0}-\eqref{eq:Ps}, both close to the bifurcation threshold and in the post-buckling regime. Readers mainly interested in results are suggested to go directly to \cref{subsec:results_num}.

\subsection{Weak formulation and finite-element approximation}
Since the problem we are studying is axisymmetric, we can simplify our analysis by reducing it to the rectangular domain $\Sigma=\left(0,L\right)\times\left(0,R_{0}\right)$ placed in the $\left(Z,\,R\right)$ plane, where $L=2\pi/\left(k_\text{cr}\lambda\right)$ is the critical wavelength of the perturbation. The fully 3D solution can be reconstructed by symmetry from the 2D solution on this cylindrical section.

The upper boundary side, i.e. $\Gamma_{4}=\{\left(Z, \,
R\right)\in\Sigma:0\leq Z\leq L,\ R=R_{0}\}$, represents the free surface of the cylinder, where the pre-stretched elastic surface is present. The following Dirichlet boundary conditions are imposed on the remaining edges
\[
\begin{aligned}
    u_{Z} & =0                       &  & \text{on }\Gamma_{1}=\{\left(Z, \, R\right)\in\Sigma:Z=0,\ 0\leq R\leq R_{0}\}, \\
    u_{Z} & =\left(\lambda-1\right)L &  & \text{on }\Gamma_{2}=\{\left(Z, \, R\right)\in\Sigma:Z=L,\ 0\leq R\leq R_{0}\},\\
    u_{R} & =0                       &  & \text{on }\Gamma_{3}=\{\left(Z, \, R\right)\in\Sigma:R=0,\ 0\leq Z\leq L\}.
\end{aligned}
\]
while homogeneous Neumann conditions are assumed for the remaining component of the traction.
In particular, on $\Gamma_1$ and $\Gamma_2$ roller-type boundary conditions are imposed: the axial displacement is prescribed to enforce the mean stretch $\lambda$, while the radial displacement is left free under zero traction. These conditions effectively act as symmetry boundaries, enabling the modelling of a segment of an infinite, uniformly stretched cylinder.

A small imperfection is applied to the mesh close to $\Gamma_{4}$ to initiate the mechanical instability. In order to follow the bifurcated branch, we employ an arclength continuation algorithm \citep{Seydel}, briefly reviewed in the following.

Let us consider the system of parametrized equations stated in an abstract setting
\begin{equation}\label{eq:f(y_eta)=0}
    \vect{f}\left(\vect{y},\,\eta\right)=\vect{0},
\end{equation}
where $\vect{y}$ represents the state of a (physical) system and $\eta$ is the control parameter of the bifurcation problem. In our application, $\vect{y}=\left(\bm{u},\,p\right)$ and $\eta$ is one of the dimensionless parameters $\widehat{\mu}, \lambda$ and $\lambda_p$.

Continuation algorithms allow us to find the region of the plane $\left(\vect{y},\eta\right)$ where \cref{eq:f(y_eta)=0} is satisfied. Specifically, we exploited a pseudo-arclength continuation algorithm, which requires the adoption of the control parameter $\eta$ as an additional unknown of the system. As a consequence, we must add a further equation to the problem. Assuming that we know that $\vect{f}(\vect{y}_j,\,\eta_j)=0$, we look for a couple $(\vect{y}_{j+1},\,\eta_{j+1})$ that satisfies \cref{eq:f(y_eta)=0} and a constraint of the form
\begin{equation}\label{eq:arclength_param}
    \|\vect{y}_{j+1}-\vect{y}_{j}\|_Y^2+\|\eta_{j+1}-\eta_j\|^2_H=\mathrm{d}s^2
\end{equation}
where $\mathrm{d}s$ is the pseudo-arclength parameter and
$\|\cdot\|_Y$ and $\|\cdot\|_H$ are suitable norms for $\vect{y}$ and $\eta$.
This equation restricts the search of $\left(\vect{y}_{j+1},\,\eta_{j+1}\right)$ to the points that are at a distance $\mathrm{d}s$ (in terms of the norms $\|\cdot\|_Y$ and $\|\cdot\|_H$) from the previous solution found at the previous step. 

The problem given by \cref{eq:f(y_eta)=0,eq:arclength_param} usually admits multiple solutions. In order to proceed along a specific path in the bifurcation diagram, a predictor-corrector method is frequently exploited. Specifically, assume that at least one solution of \cref{eq:f(y_eta)=0} can be found, say, $\left(\vect{y}_{1},\eta_1\right)$. Then, the $j{\text{-th}}$ continuation step attempts to find the solution $\left(\vect{y}_{j+1},\eta_{j+1}\right)$ starting from the previously calculated $\left(\vect{y}_{j},\eta_{j}\right)$. This process is usually split into two parts: the former is called \emph{predictor step} and denoted by $\left(\Bar{\vect{y}}_{j+1},\,\Bar{\eta}_{j+1}\right)$. It is aimed at finding a good approximation of $\left(\vect{y}_{j+1},\eta_{j+1}\right)$ without necessarily being a solution of \cref{eq:f(y_eta)=0}. The latter is named \emph{corrector step} and, starting from the output of the predictor step, will produce an effective solution of \cref{eq:f(y_eta)=0}. It usually consists in a Newton-Rapson algorithm.
\begin{equation*}
\left(\vect{y}_{j},\eta_{j}\right)\xrightarrow{\text{predictor}} \left(\Bar{\vect{y}}_{j+1},\Bar{\eta}_{j+1}\right)\xrightarrow{\text{corrector}} \left(\vect{y}_{j+1},\eta_{j+1}\right).
\end{equation*}
This induces the corrector to modify the predictor output in order to find a solution that satisfies \cref{eq:arclength_param}.

In our application, in \cref{eq:arclength_param} for the state $\vect{y} = (\bm{u},\,p)$ and the control parameter $\eta$, we adopt the $L^2$ norm over the referential domain $\mathscr{B}_0$. In the simulations presented in this work, we used a secant predictor, i.e
\begin{equation*}
    \left(\Bar{\vect{y}}_{j+1},\Bar{\eta}_{j+1}\right)=\left(\vect{y}_{j},\eta_{j}\right)+\left(\vect{y}_{j}-\vect{y}_{j-1},\eta_{j}-\eta_{j-1}\right),
\end{equation*}
meaning that the predictor is chosen on the prolongation of the segment $\left(\vect{y}_{j}-\vect{y}_{j-1},\eta_{j}-\eta_{j-1}\right)$.

To introduce the weak formulation of the two-dimensional problem, we define
the following functional spaces
\begin{align*}
    \mathcal{V}     & =\Big\{\bm{v}\in \left[H^{1}\left(\Sigma\right)\right]^{2}:v_{R}=0 \text{ on }\Gamma_{3}\text{, }v_{Z}=0 \text{ on }\Gamma_{1}\text{, }v_{Z}=\left(\lambda-1\right)L\text{ on }\Gamma_{2}\Big\}, \\
    \mathcal{V}_{0} & =\Big\{\bm{v}\in \left[H^{1}\left(\Sigma\right)\right]^{2}:v_{R}=0 \text{ on }\Gamma_{2}\cup\Gamma_{3},\, v_{Z}=0 \text{ on }\Gamma_{1}\Big\},                                              \\
    \mathcal{Q}     & =L^{2}\left(\Sigma\right).
\end{align*}
Specifically, $\mathcal{V}$ and $\mathcal{V}_{0}$ represent the spaces where
the trial and test functions for the displacement, respectively, will be sought,
while $\mathcal{Q}$ is the functional space for trial and test functions
associated to the pressure field. The parameter space is simply $\mathbb{R}$. To sum up, the weak formulation of the arclength problem reads: find $\left(\bm u_{j+1},\,p_{j+1},\,\eta_{j+1}\right)\in \mathcal{V}\times \mathcal{Q}\times\mathbb{R}$ such that
\begin{subequations}
    \begin{align}
            & \int_{\Sigma}\tens{P}_{j+1}:\nabla\bm{v}\ \mathrm{d}A_{0}+\int_{\Gamma_4}\tens{P}_{s\,j+1}:\nabla_{s}\bm{v}\ \mathrm{d}\ell_{0}=0                                                                                                                   &  & \forall \ \bm{v}\in \mathcal{V}_{0},\label{eq:weak_1} \\
            & \int_{\Sigma}\left(\det \tens{F}_{j+1}- 1\right) q\ \mathrm{d}A_{0}=0                                                                                                                                                                                 &  & \forall \ q\in \mathcal{Q},\label{eq:weak_2}          \\
            & \delta\eta\left(\int_{\Sigma}\left(|\vect{u}_{j+1}-\vect{u}_{j}|^{2}+\left(p_{j+1}-p_{j}\right)^{2}\right) 2 R\, \mathrm{d}A_{0}+\left(\eta_{j+1}-\eta_{j}\right)^{2}|\Sigma|R_{0}\right) = \delta\eta\,\mathrm{d}s^{2}|\Sigma|R_{0} &  & \forall \ \delta\eta\in \mathbb{R},\label{eq:weak_3}
    \end{align}
\end{subequations}
where the Piola-Kirchhoff tensors and the deformation gradient tensor are evaluated
at $\left(\vect{u},p\right)=\left(\vect{u}_{j+1},p_{j+1}\right)$.

We introduce the discretization of the problem by defining an isotropic triangulation $\mathscr{T}=\bigcup_{i=1}^{n_e}\mathcal{K}_{i}$ over $\Sigma$, where, for every $i$, $\mathcal{K}_{i}$ is a triangle in $\Sigma$ and $n_{e}$ is the total number
of triangles.
In each simulation, we discretize the computational domain using a structured mesh with 30 elements along the radial direction.
\Cref{eq:weak_1,eq:weak_2,eq:weak_3} are discretized using a stable
pair of continuous finite dimensional spaces belonging to the family of Taylor-Hood elements
\citep{quarteroni}. 
In particular, we use $\mathbb{P}^{2}$ elements for the
displacement field and $\mathbb{P}^{1}$ elements for the pressure over each
triangle $\mathcal{K}_{i}$ in $\mathscr{T}$, where $\mathbb{P}^r$ denotes the space of polynomials of degree $r$ over the triangle $\mathcal{K}_{i}$ that are continuous over the physical domain.
The proposed numerical scheme is implemented in Python using the finite element computing platform \texttt{FEniCS} \citep{fenics} and
library \texttt{BiFEniCS} that allows to implement the continuation algorithm \citep{bifenics}. 

\subsection{Results of the numerical simulation}\label{subsec:results_num}

\begin{figure}[t!]
    \centering
    \includegraphics[width=0.8\textwidth]{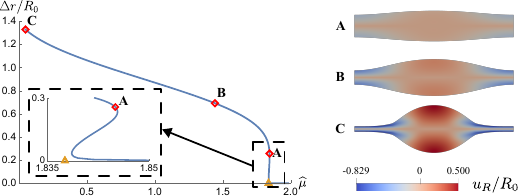}
    \caption{(Left) Bifurcation diagram showing the dimensionless beading amplitude
    $\Delta r/R_{0}$ (see Eq.~\eqref{eq:amplitude}) versus the dimensionless shear modulus $\widehat{\mu}$. Here, the axial strain $\lambda$, the dimensionless surface extensibility $\widehat{\Lambda}_{s}$, and the pre-stretch parameter $\lambda_p$ are set to 1.4, 40 and 0.8, respectively. The cylindrical configuration becomes unstable when $\widehat{\mu}$ decreases below the critical threshold $\widehat{\mu}_\text{cr}$. The orange triangle denotes the theoretical stability threshold obtained with the linear stability analysis. (Right) Buckled morphology obtained for the three values of $\widehat{\mu}$, corresponding to the three points A, B and C reported in the bifurcation diagram.}
    \label{fig:bif_diag_mu}
\end{figure}

We start by studying the bifurcations induced in the fully nonlinear problem by variations of the dimensionless shear modulus $\widehat{\mu}$. The bifurcation diagram, reported in \cref{fig:bif_diag_mu}, shows the amplitude of the beading pattern $\Delta r$ versus the dimensionless shear modulus $\widehat{\mu}$, where
\begin{equation}
\label{eq:amplitude}
    \Delta r=\max_{Z\in\left[0,\,2\pi R_0/\left(\lambda \widehat{k}_\text{cr}\right)\right]}
    r\left(R_{0},\,Z\right)-\min_{Z\in\left[0,\,2\pi R_0/\left(\lambda
    \widehat{k}_\text{cr}\right)\right]}r\left(R_{0},\,Z\right).
\end{equation}
The plot in \cref{fig:bif_diag_mu} shows that the cylinder undergoes a subcritical pitchfork bifurcation when it reaches the marginal stability threshold predicted by the linear stability analysis.
In particular, we notice that the cylindrical configuration remains stable as long as $\widehat
{\mu}$ is greater than the bifurcation threshold. Nearby the marginal stability threshold, the buckled morphology reproduces the sinusoidal pattern of the linear stability analysis. In the nonlinear regime, the system exhibits the formation of bulges spaced with long, extremely thinned regions: this phenomenon is caused by an increasingly localized beading pattern due to a progressive decrease in $\widehat{\mu}$. Structures like these are typically observed in damaged axons, where a similar morphological instability occurs when the degraded cytoskeleton is squeezed by the action of the surrounding actin cortex \citep{Datar_et_al,Riccobelli,fu2021necking,dehghany2024osmotically}.

\begin{figure}[t!]
    \centering
    \includegraphics[width=0.8\textwidth]{
        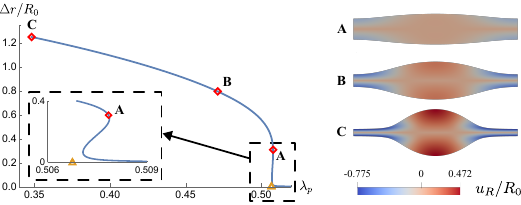
    }
    \caption{(Left) Bifurcation diagram showing the dimensionless beading amplitude
    $\Delta r/R_{0}$ (see Eq.~\eqref{eq:amplitude}) versus the pre-stretch parameter $\lambda_{p}$. Here, the axial strain $\lambda$,  the dimensionless surface extensibility $\widehat{\Lambda}_{s}$ and  the dimensionless shear modulus $\widehat{\mu}$ are set to 1.4, 40 and 20.5, respectively. The cylindrical configuration becomes unstable when $\lambda_{p}$ decreases below the critical threshold $\lambda_{p}^{cr}$. The orange triangle denotes the theoretical stability threshold obtained with the linear stability analysis. (Right) Buckled morphology obtained for the three values of $\lambda_p$, corresponding to the three points A, B and C reported in the bifurcation diagram.}
    \label{fig:bif_diag_la}
\end{figure}

Similarly, we study the behaviour of the cylinder subjected to progressive variations of the pre-stretch parameter $\lambda_{p}$, simulating the increase of surface tension induced by the change of the medium surrounding the cylinder, as in the experiments of \cite{Mora_2013}. In \cref{fig:bif_diag_la} we show the dimensionless beading
amplitude against $\lambda_p$.
Similarly to $\widehat{\mu}$, the bifurcation diagram shows a subcritical pitchfork bifurcation that stabilizes in the nonlinear regime. %
We deduce that the morphology remains stable in a cylindrical shape when $\lambda_p$ is sufficiently high, i.e. for small surface pre-stretch. Close to the bifurcation value, the cylindrical profile begins to deform, leading to a similar periodic beading as in \cref{fig:bif_diag_mu}. Therefore, a decrease of $\lambda_p$ corresponds to a more and more pronounced separation between bulges and thinned regions. We recall that the relation between $\lambda_p$ and surface tension is discussed in \Cref{rem:st_lambdap}, and smaller values of $\lambda_p$ are associated with an increased surface stress.

\begin{figure}[t!]
    \centering
    \subfloat[$\lambda_p=0.7$.\label{fig:k_lambda_bif_07}]{ \includegraphics[width=0.45\textwidth]{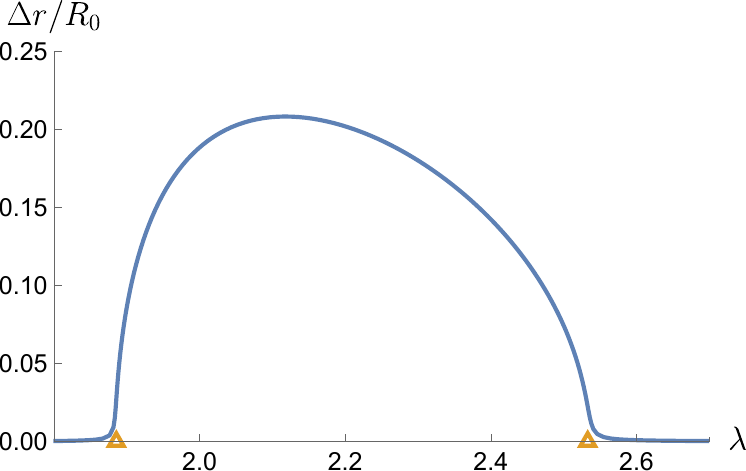} }
    \quad \subfloat[$\lambda_p=0.6$.\label{fig:k_lambda_bif_06}]{ \includegraphics[width=0.45\textwidth]{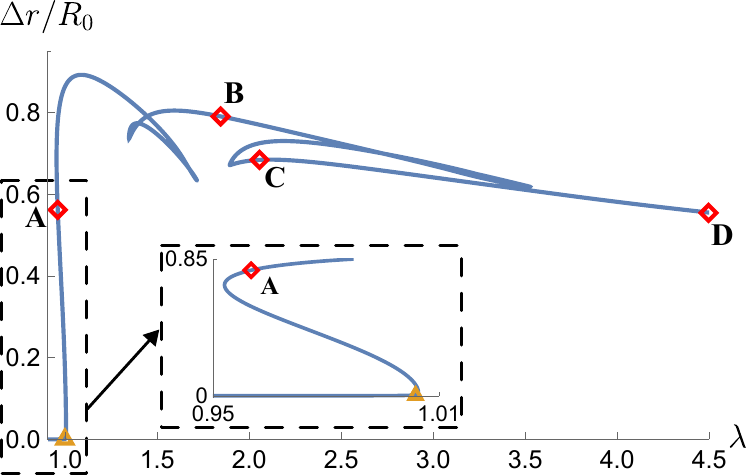} }
    \\
    \subfloat[Buckled morphologies obtained for different values of $\lambda$ and $\lambda_p=0.6$.\label{fig:lambda_configurations}]{ \includegraphics[width=0.9\textwidth]{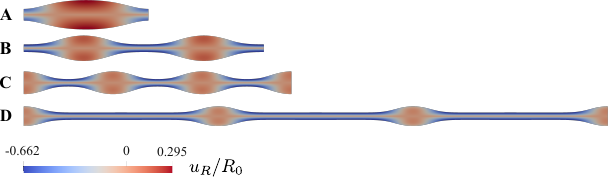} }
    \caption[]{(Top) Bifurcation diagrams showing the dimensionless beading amplitude $\Delta r/R_{0}$ (see Eq.~\eqref{eq:amplitude}) versus the axial stretch $\lambda$. Here, the dimensionless shear modulus $\widehat{\mu}$ and the dimensionless surface extensibility $\widehat{\Lambda}_s$ are set to 0.8 and 10, respectively, while the pre-stretch parameter $\lambda_p$ is equal to 0.7 (a) and 0.6 (b). The orange triangles denote the marginal stability thresholds obtained with the linear stability analysis. (Bottom) Beaded morphology of the buckled cylinder for $\lambda_p=0.6$, corresponding to the points A, B, C, and D reported in the bifurcation diagram of panel (b).}
    \label{fig:linear_stability_analysis_k_lambda}
\end{figure}

Finally, we perform the post-buckling simulations by continuously varying the axial strain $\lambda$. We notice different behaviors depending on $\lambda_p$. As shown in \cref{fig:linear_stability_analysis_k_lambda}, for moderate pre-stretch, we observe a supercritical bifurcation. If the body is further axially stretched, the system returns to the unbuckled cylindrical configuration. This possibility was also explored by \cite{taffetani_ciarletta}.  
On the other hand, for smaller values of $\lambda_p$, a subcritical pitchfork bifurcation occurs (\cref{fig:k_lambda_bif_06}). In the nonlinear regime, if the cylinder is further stretched, the instability is not suppressed, as in the previous case. Instead, a sequence of period-halving secondary bifurcations appears. %

\section{Final remarks}
\label{sec:conclusion} 

Surface phenomena in soft elastic media have attracted significant attention due to their relevance in material science, biophysics, and engineering. In this work, we have incorporated elastic effects into surface tension to effectively describe the Plateau-Rayleigh instability in solids.

Specifically, we have adopted the theory of material surfaces proposed by \cite{Gurtin_Murdoch} to account for elastic pre-stretch. Through linear stability analysis, we have shown that a cylindrical solid can undergo mechanical instability when surface energy is sufficiently high relative to bulk elastic energy. In particular, a mechanical instability can be triggered by
\begin{itemize}
    \item decreasing bulk elastic stiffness,
    \item applying axial stretch,
    \item increasing surface tension.
\end{itemize}

All these scenarios are thoroughly analysed in \cref{sec:analysis_of_the_model}, where we prove that the critical wavenumber falls within the interval $0.5 R_0<k_{cr}<0.7R_0$, $R_0$ being the reference radius. This range qualitatively aligns with experimental findings from \citep{matsuo1992patterns,PhysRevLett.105.214301} and has not been captured by previous mathematical models, highlighting the crucial role of surface tension elasticity in reproducing this phenomenon. The only exception is provided by \cite{TAFFETANI2024105606}, which incorporated bending elastic energy alongside a constant surface tension, suggesting that elasticity may play a role in triggering a periodic pattern. We remark that the boundary layer generated by capillary effects is relatively small, meaning that stretching energy should dominate over bending terms, thereby justifying the approach proposed in this paper. Nevertheless, a promising avenue for exploring the interplay between stretching and bending is offered by the theoretical framework proposed by \cite{tomassetti2024coordinate}.

The postbuckling behavior is investigated using a finite element approximation combined with a continuation algorithm, with results presented in \cref{sec:numerical_simulations}. Depending on the parameter values, the resulting buckled states stem from either subcritical or supercritical pitchfork bifurcations. Interestingly, when buckling is induced by axial stretching, we observe either a reversal of the bifurcation, where the system returns to a straight cylindrical configuration after transient buckling at higher stretches, or a sequence of period-halving bifurcations.

The potential role of secondary bifurcations in wavelength selection has been conjectured in \cite{PhysRevLett.105.214301} and our study confirms that they play an important role far from the bifurcation threshold, leading to a cascade of period-halving events.

Our work offers a possible explanation for the onset and development of beading in soft elastic filaments and underscores the importance of incorporating strain-dependent surface tension at the interface of soft solids. Indeed, key characteristics of the elastic Rayleigh-Plateau instability, such as the finite wavelength, cannot be captured without accounting for this elastic dependence.  
Future efforts will be devoted to understanding the behaviour of this system when subjected to structural damage. Hydrogel filaments have been observed to develop voids following mechanical instability \citep{matsuo1992patterns}, a phenomenon reminiscent of elastic cavitation. Studying this system may shed light on the structural damage occurring in neurons during neurodegenerative diseases, where similar beaded structures are observed in axons \citep{PhysRevLett.96.048104, Datar_et_al}. Moreover, an experimental characterization of material surface parameters would be of interest, as has been done for linearly elastic material surfaces in \cite{Heyden_2022}.

\section*{Acknowledgments} 
This work has been partially supported by INdAM through the project \emph{MATH-FRAC: MATHematical modelling of FRACture in nonlinear elastic materials} and by PRIN 2022 project \emph{Mathematical models for viscoelastic biological matter}, Prot. 202249PF73 – Funded by European Union - Next Generation EU - Italian Recovery and Resilience Plan (PNRR) - M4C1\_CUP D53D23005610001. Financial support from the Italian Ministry of University and Research (MIUR) through the grant "Dipartimenti di Eccellenza 2023-2027 (Mathematics Area)” is gratefully acknowledged. The authors are members of GNFM -- INdAM (National Group of Mathematical Physics).

\bibliographystyle{apalike}
\bibliography{bibliography.bib}

\appendix
\section{Differential operators in curvilinear coordinates}
\label{appendix:geometry}

In this appendix we suppose that indices $i$ and $j$ run from 1 to 3, while
$\alpha$ and $\beta$ run from 1 to 2. We denote by
$\mathcal{E}=\{\vect{e}_{1},\,\vect{e}_{2},\,\vect{e}_{3}\}$ the canonical basis of
$\mathbb{R}^{3}$. Let $\mathcal{G}=\{\bm{g}_{1},\,\bm{g}_{2},\,\bm{g}_{3}\}$ be an arbitrary basis in $\mathcal{E}$.
We say that $\mathcal{G}^{*}=\{\bm{g}^{1},\,\bm{g}^{2},\,\bm{g}^{3}\}$ is the dual basis to
$\mathcal{G}$ if and only if $\bm{g}_{i}\cdot\bm{g}^{j}=\delta_{i}^{j}$ for
every $i$ and $j$, where $\delta_i^j$ denotes the Kronecker delta. Duality is a reflexive property and, given the basis $\mathcal{G}$, there always exists its dual $\mathcal{G}
^{*}$. It is straightforward to see that $\mathcal{E}^{*}=
\mathcal{E}$.

Given an arbitrary vector $\bm{v}$, we denote by $\{v_{1},\,v_2,\,v_3\}$ its \emph{covariant
components} and by $\{v^{1},\,v^2,\,v^3\}$ its \emph{contravariant components}. Let us assume
that $\{\bm{g}^{1},\,\bm{g}^{2},\,\bm{g}^{3}\}$ be the basis of a generic curvilinear coordinates system
$\{\xi^{1},\,\xi^{2},\,\xi^{3}\}$ and let $d\bm{r}$ denote an arbitrary infinitesimal vector
expressed in terms of the Cartesian coordinates $\{x^{1},\,x^{2},\,x^{3}\}$. It is
always possible to express $d\bm{r}$ in terms of the curvilinear coordinates
$\{\xi^{1},\,\xi^2,\,\xi^3\}$ through the linear, invertible relation $\bm{\alpha}$ between
the two coordinates systems, that is $\vect{e}_{i}=\alpha_{i}^{j}\bm{g}^{j}$.
The relation between $\bm{g}_{i}$, $\bm{g}^{i}$ and $d\bm{r}$ are expressed
as
\begin{align*}
    \bm{g}_{i}=\frac{\partial\bm{r}}{\partial\xi^{i}}, &  & \bm{g}^{i}=\frac{\partial\xi^{i}}{\partial\bm{r}}.
\end{align*}

Moreover, we denote by $g^{ij}$ the map from the contravariant basis to the
covariant one and by $g_{ij}$ the mapping from the covariant basis to the
contravariant one, namely
\begin{align*}
        & \bm{g}^{i}=g^{ij}\bm{g}_{j}, \\
        & \bm{g}_{i}=g_{ij}\bm{g}^{j}.
\end{align*}
The coefficients of these two maps are called covariant and contravariant metric
coefficients, respectively. They are defined as $g_{ij}=\bm{g}_{i}\cdot\bm{g}
_{j}$ and $g^{ij}=\bm{g}^{i}\cdot\bm{g}^{j}$ and are such that $\left[g_{ij}\right
]=\left[g^{ij}\right]^{-1}$.\\ We can introduce the gradient and divergence operators
using the general curvilinear coordinates introduced above \citep{javili_et_al}
\begin{subequations}
    \label{eq:grad_div_det_bulk}
    \begin{align}
            & \grad\,\{\cdot\}=\frac{\partial\{\cdot\}}{\partial\xi^{i}}\otimes\bm{g}_{i}, \label{eq:grad}                         \\
            & \diver\,\{\cdot\}=\frac{\partial\{\cdot\}}{\partial\xi^{i}}\cdot\bm{g}_{i}=\grad\,\{\cdot\}:\tens{I}, \label{eq:div}
    \end{align}
\end{subequations}
where $\tens{I}$ is the identity tensor in $\mathbb{R}^{3}$.

Let us consider a regular surface $\mathcal{S}$ in the current configuration.
Being
$\mathscr{P}=(\widehat{\xi}^{1},\,\widehat{\xi}^{2})\subset\mathbb{R}^{2}$, let
$\bm{\xi}:\mathscr{P}\rightarrow\mathcal{S}$ be a parametrization of the surface.
In analogy to the procedure derived for the bulk, we define the covariant and
contravariant surface basis vectors for the curvilinear coordinates as
\begin{align*}
    \widehat{\bm{g}}_{\alpha}=\frac{\partial\bm{r}}{\partial\widehat{\xi}^{\alpha}}, &  & \widehat{\bm{g}}^{\alpha}=\frac{\partial\widehat{\xi}^{\alpha}}{\partial\bm{r}}.
\end{align*}
As for the bulk, there exists an invertible relation between co- and contravariant
surface basis vectors, that is
\begin{align*}
        & \widehat{\bm{g}}^{\alpha}=\widehat{g}^{\alpha\beta}\widehat{\bm{g}}_{\beta}, \\
        & \widehat{\bm{g}}_{\alpha}=\widehat{g}_{\alpha\beta}\widehat{\bm{g}}^{\beta},
\end{align*}
where
$\left[\widehat{g}_{\alpha\beta}\right]=\left[\widehat{g}^{\alpha\beta}\right
]^{-1}$
with
$\widehat{g}_{\alpha\beta}=\widehat{\bm{g}}_{\alpha}\cdot\widehat{\bm{g}}_{\beta}$
and
$\widehat{g}^{\alpha\beta}=\widehat{\bm{g}}^{\alpha}\cdot\widehat{\bm{g}}^{\beta}$.
Now, it is possible to define the contra- and covariant base vectors that are
normal to the surface $\mathcal{S}$ as $\widehat{\bm{g}}^{3}\coloneqq\widehat
{\bm{g}}^{1}\wedge\widehat{\bm{g}}^{2}$ and $\widehat{\bm{g}}_{3}=\left[\widehat
{g}^{33}\right]^{-1}\widehat{\bm{g}}^{3}$ in such a way that
\begin{equation*}
    \widehat{\bm{g}}_{3}\cdot\widehat{\bm{g}}^{3}=1,
\end{equation*}
coherently with the definition of dual basis. As a consequence, the normal unit
vector to the surface is
\begin{equation}
    \label{eq:normal_unit_vector}\bm{n}=\frac{\widehat{\bm{g}}_{3}}{\lvert\widehat{\bm{g}}_{3}\rvert}
    =\frac{\widehat{\bm{g}}^{3}}{\lvert\widehat{\bm{g}}^{3}\rvert},
\end{equation}
where the last equality holds since $\widehat{\bm{g}}_{3}$ and $\widehat{\bm{g}}
^{3}$ are parallel. The surface identity tensor in the current configuration is
defined as
\begin{equation}
    \label{eq:i_s_appendix}\tens{H}_{s}=\tens{I}-\widehat{\bm{g}}_{3}\otimes\widehat
    {\bm{g}}^{3}=\tens{I}-\bm{n}\otimes\bm{n},
\end{equation}
that is exactly \cref{eq:relation i/I}.\\

We can finally define the surface gradient, divergence and determinant
operators as \citep{javili_et_al}
\begin{subequations}
    \label{eq:grad_div_det_surface}
    \begin{align}
            & \grad_{s}\,\{\cdot\}=\frac{\partial\{\cdot\}}{\partial\widehat{\xi}^{\alpha}}\otimes\widehat{\bm{g}}_{\alpha}, \label{eq:grad_sup}                                                                                               \\
            & \diver_{s}\,\{\cdot\}=\frac{\partial\{\cdot\}}{\partial\widehat{\xi}^{\alpha}}\cdot\widehat{\bm{g}}_{\alpha}=\grad_{s}\,\{\cdot\}:\tens{H}_{s}, \label{eq:div_sup}                                                               \\
            & \text{det}_{s}\,\{\cdot\}=\frac{\lvert\left[\{\cdot\}\cdot\widehat{\bm{g}}_{1}\right]\wedge\left[\{\cdot\}\cdot\widehat{\bm{g}}_{2}\right]\rvert}{\lvert\widehat{\bm{g}}_{1}\wedge\widehat{\bm{g}}_{2}\rvert}.\label{eq:det_sup}
    \end{align}
\end{subequations}
To conclude, the \emph{surface divergence theorem} holds: let $\sigma$ be a
regular subsurface of $\mathcal{S}$ with a smooth boundary $\partial\sigma$
and let $\bm{m}$ be the outward unit normal to $\partial\sigma$. Then
\begin{equation}
    \label{thm:surface_div_thm}\int_{\sigma}\bm{T}\cdot\bm{m}=\int_{\partial\sigma}
    \textnormal{div}_{s}\, \bm{T},
\end{equation}
for every tangent field $\bm{T}$ to $\sigma$. In the computations and
definitions of this appendix we have exploited the notation we have used for
material bodies and surfaces in the current configuration. Nonetheless, all these
results are general and can be applied to every framework. However,
particular attention should be devoted to the surface identity tensors: if on
one hand the reference and actual bulk identities are invariant and equal \citep{javili_et_al},
the same cannot be stated for the surface ones. Indeed, while the surface actual
identity has been defined in \cref{eq:i_s_appendix}, given a surface
$\mathcal{S}_{0}$ in reference configuration, we have
\begin{equation*}
    \tens{I}_{s}=\tens{I}-\vect{N}\otimes\vect{N},
\end{equation*}
where $\vect{N}$ is the normal unit vector to $\mathcal{S}_{0}$. Since, in general,
$\mathcal{S}_{0}\neq\mathcal{S}$, also $\vect{N}\neq\bm{n}$. Thus, we conclude
that in general $\tens{I}_{s}\neq\tens{H}_{s}$.

\section{Coefficients of the matrix \texorpdfstring{$\tens{M}$}{M}}
\label{appendix:matrix} In this appendix we provide the explicit expressions
of the components of the $2\times2$ matrix $\tens{M}$ arising from the imposition
of the boundary condition \cref{eq:incr_div_sP_s=PN}. Since the expression
for $U(r)$ depends on the value of $\lambda$ (see \cref{eq:U(R)_lambda!=1,eq:U(R)_lambda=1})
we distinguish the two cases.\\

When $\lambda=1$ we obtain
\begin{subequations}
    \begin{align*}
        \begin{split}M_{11}&=J_{1}\left(k R_{0}\right) \left(2 \lambda_{p}^{4}\left(-2 \mu R_{0}+\Lambda_{s}+2 \mu_{s}\right)-k^{2}\left(\lambda_{p}^{2}-1\right) R_{0}^{2}\left(\lambda_{p}^{2}\Lambda_{s}+\Lambda_{s}+2 \lambda_{p}^{2}\mu_{s}\right)\right)+ \\&\quad + k R_{0}J_{0}\left(k R_{0}\right) \left(2\lambda_{p}^{2}\left(2 \lambda_{p}^{2}\mu R_{0}-\lambda_{p}^{2}\mu_{s}+\mu_{s}\right)-\left(\lambda_{p}^{4}+1\right)\Lambda_{s}\right),\end{split} \\
        \begin{split}M_{12}&=-R_{0}\big(J_{0}\left(k R_{0}\right) \left(\Lambda_{s}\left(k^{2}\left(\lambda_{p}^{4}-1\right) R_{0}^{2}+2\right)+2 \lambda_{p}^{2}\mu_{s}\left(k^{2}\left(\lambda_{p}^{2}-1\right) R_{0}^{2}-2\right)\right)+ \\&\quad +k R_{0}J_{1}\left(k R_{0}\right) \left(-4 \lambda_{p}^{4}\mu R_{0}+\lambda_{p}^{4}\Lambda_{s}+\Lambda_{s}+2 \lambda_{p}^{4}\mu_{s}-2 \lambda_{p}^{2}\mu_{s}\right)\big),\end{split}                            \\
        \begin{split}M_{21}&=k \big(k R_{0}J_{0}\left(k R_{0}\right) \left(\left(\lambda_{p}^{4}+1\right) \Lambda_{s}+2 \left(\lambda_{p}^{2}+1\right) \lambda_{p}^{2}\mu_{s}\right)+ \\&\quad - J_{1}\left(k R_{0}\right) \left(-4 \lambda_{p}^{4}\mu R_{0}+\lambda_{p}^{4}\Lambda_{s}+\Lambda_{s}+2 \lambda_{p}^{4}\mu_{s}-2 \lambda_{p}^{2}\mu_{s}\right)\big),\end{split}                                                                                         \\
        \begin{split}M_{22}&=R_{0}\big(J_{1}\left(k R_{0}\right) \left(k^{2}R_{0}\left(\lambda_{p}^{4}\Lambda_{s}+\Lambda_{s}+2 \lambda_{p}^{4}\mu_{s}+2 \lambda_{p}^{2}\mu_{s}\right)+4 \lambda_{p}^{4}\mu \right)+ \\&\quad + k J_{0}\left(k R_{0}\right) \left(4 \lambda_{p}^{4}\mu R_{0}+\lambda_{p}^{4}\Lambda_{s}+\Lambda_{s}+2 \lambda_{p}^{4}\mu_{s}+6 \lambda_{p}^{2}\mu_{s}\right)\big).\end{split}
    \end{align*}
\end{subequations}

If $\lambda\neq 1$, we get
\begin{subequations}
    \begin{align*}
        \begin{split}M_{11}&=\frac{1}{\lambda^{3}}\Big(-2 J_{1}\left(k \lambda R_{0}\right) \left(k^{2}R_{0}^{2}\left(2 \lambda^{2}\lambda_{p}^{2}\mu_{s}+\lambda \Lambda_{s}-\left(\lambda_{p}^{4}\left(\Lambda_{s}+2 \mu_{s}\right)\right)\right)-\lambda^{2}\Lambda_{s}+\lambda \lambda_{p}^{4}\left(\Lambda_{s}+2 \mu_{s}\right)+2\lambda_{p}^{2}\mu_{s}\right)+ \\&\qquad\quad + k \lambda R_{0}J_{0}\left(k \lambda R_{0}\right) \left(-2 \lambda_{p}^{2}\left(2 \lambda_{p}^{2}\mu R_{0}+\mu_{s}\right)+\lambda^{2}\Lambda_{s}+\lambda \lambda_{p}^{4}\left(\Lambda_{s}+2 \mu_{s}\right)\right)+ \\&\qquad\quad + k \lambda R_{0}J_{2}\left(k\lambda R_{0}\right) \left(-2 \lambda_{p}^{2}\left(2 \lambda_{p}^{2}\mu R_{0}+\mu_{s}\right)+\lambda^{2}\Lambda_{s}+\lambda \lambda_{p}^{4}\left(\Lambda_{s}+2 \mu_{s}\right)\right)\Big),\end{split}                                                                                                                                                                                                \\
        \begin{split}M_{12}&=\frac{1}{\lambda^{2}}\Big(2 \sqrt{\lambda }J_{1}\left(\frac{k R_{0}}{\sqrt{\lambda}}\right) \big(k^{2}R_{0}^{2}\left(2 \lambda^{2}\lambda_{p}^{2}\mu_{s}+\lambda \Lambda_{s}-\left(\lambda_{p}^{4}\left(\Lambda_{s}+2 \mu_{s}\right)\right)\right)+ \\&\qquad\quad +2 \lambda^{3}\lambda_{p}^{4}\mu R_{0}-2 \lambda_{p}^{4}\mu R_{0}-\lambda^{2}\Lambda_{s}+\lambda \lambda_{p}^{4}\left(\Lambda_{s}+2 \mu_{s}\right)+2 \lambda_{p}^{2}\mu_{s}\big)+ \\&\qquad\quad +k R_{0}J_{0}\left(\frac{k R_{0}}{\sqrt{\lambda}}\right)\left(2 \lambda^{3}\lambda_{p}^{4}\mu R_{0}+2 \lambda_{p}^{2}\left(\lambda_{p}^{2}\mu R_{0}+\mu_{s}\right)-\lambda^{2}\Lambda_{s}-\lambda \lambda_{p}^{4}\left(\Lambda_{s}+2 \mu_{s}\right)\right)+ \\&\qquad\quad +k R_{0}J_{2}\left(\frac{k R_{0}}{\sqrt{\lambda}}\right) \left(2 \lambda^{3}\lambda_{p}^{4}\mu R_{0}+2 \lambda_{p}^{2}\left(\lambda_{p}^{2}\mu R_{0}+\mu_{s}\right)-\lambda^{2}\Lambda_{s}-\lambda \lambda_{p}^{4}\left(\Lambda_{s}+2 \mu_{s}\right)\right)\Big),\end{split} \\
        \begin{split}M_{21}&=\frac{k}{\lambda^{3/2}}\Big(4 \left(\lambda^{3}+1\right) \lambda_{p}^{2}J_{1}\left(k \lambda R_{0}\right) \left(\lambda_{p}^{2}\mu R_{0}+\mu_{s}\right)+k \lambda^{2}R_{0}J_{0}\left(k \lambda R_{0}\right) \left(2 \lambda^{2}\lambda_{p}^{2}\mu_{s}+\lambda \Lambda_{s}+\lambda_{p}^{4}\left(\Lambda_{s}+2\mu_{s}\right)\right)+ \\&\qquad\quad +k \lambda^{2}R_{0}J_{2}\left(k \lambda R_{0}\right) \left(2 \lambda^{2}\lambda_{p}^{2}\mu_{s}+\lambda \Lambda_{s}+\lambda_{p}^{4}\left(\Lambda_{s}+2 \mu_{s}\right)\right)\Big),\end{split}                                                                                                                                                                                                                                                                                                                                                                                                                                                                              \\
        \begin{split}M_{22}&=\frac{k}{\lambda^{5/2}}\Big(4 \lambda \lambda_{p}^{2}J_{1}\left(\frac{k R_{0}}{\sqrt{\lambda}}\right) \left(2 \lambda_{p}^{2}\mu R_{0}+\lambda^{3}\mu_{s}+\mu_{s}\right)+ \\&\qquad\quad +k \lambda^{3/2}R_{0}J_{0}\left(\frac{k R_{0}}{\sqrt{\lambda}}\right) \left(2 \lambda^{2}\lambda_{p}^{2}\mu_{s}+\lambda \Lambda_{s}+\lambda_{p}^{4}\left(\Lambda_{s}+2 \mu_{s}\right)\right)+ \\&\qquad\quad + k \lambda^{3/2}R_{0}J_{2}\left(\frac{k R_{0}}{\sqrt{\lambda}}\right) \left(2 \lambda^{2}\lambda_{p}^{2}\mu_{s}+\lambda \Lambda_{s}+\lambda_{p}^{4}\left(\Lambda_{s}+2 \mu_{s}\right)\right)\Big).\end{split}
    \end{align*}
\end{subequations}
\end{document}